\documentclass{article}
\usepackage{amssymb}
\usepackage{slashbox}
\usepackage[pdftex]{graphicx}
\usepackage[utf8]{inputenc}
\usepackage{amsmath} 
\usepackage[T1]{fontenc}

\title{ -Deformed logarithms.\\-Associated entropic divergences.\\
 -Applications to linear inverse problems.\\
- Inversion algorithms.}
\author{ Henri Lantéri}

\begin{document}
\maketitle
\tableofcontents
\section{Introduction and Contents}
This work belongs to the framework of inverse problems with linear model. The resolution of this type of problem consists in minimizing (possibly under constraints) a function of discrepancy between the measurements and a physical model of the considered phenomenon.\\
In the context of image deconvolution problems, for example, the constraints considered are typically the non-negativity constraint and the constraint on the sum of the total intensity of the reconstructed image.\\ 
This approach implies, on the one hand, the definition of a discrepancy function and, on the other hand, the implementation of an algorithmic method for the minimization of such a function under constraints.\\
These two aspects have been developed in a previous work \cite{lanteri2019}, \cite{lanteri2020}.\\
In chapter 3 of the above-mentioned work, divergences based on the most classical forms of entropy were analysed.\\
We will here extend this study to divergence functions based on forms of entropy developed in the field of Statistical Physics.\\
These various forms of entropy are based on ``\textit{the deformed algebra}'' \cite{kaniadakis2002} and in particularly on the notion of ``\textit{Deformed Logarithm}'' \cite{naudts2002} \cite{kaniadakis2002}.\\
It is obvious that the analysis of the properties of such entropies is not in the scope of our work and that we will only use the results found in the literature.\\
As a consequence, this document is organised as follows.\\
In sections 2 and 3, we give some brief reminders about the inverse problems and about entropy.\\
In sections 4 to 9, we will review the main forms of entropy used in statistical physics and we will analyse the divergences built on such functions.\\
In section 10, we consider other definitions of entropy based on the concepts of generalized derivatives. This allows us to recover the different forms of entropy proposed in the previous sections, but more importantly it allows us to highlight a very general form of entropy that synthesises all the previous ones.\\
In section 11, we analyse the divergences constructed on the basis of this general form of entropy.\\
Section 12 is dedicated to some forms of entropy that fall outside the general framework previously described in Section 10 and the divergences associated with these particular entropies are analysed.\\
In Section 13, we consider scale-invariant divergences based on the general entropy presented in Section 11; the different forms of divergences are considered and the special case of Tsallis entropy is analysed in more details.\\
Section 14 is dedicated to the algorithmic aspect.\\
After a brief review of the SGM method more widely developed in \cite{lanteri2019}, \cite{lanteri2020}, we describe the corresponding algorithms allowing the minimization of general divergences in the case of a non-negativity constraint, and then, in the case of an additional constraint on the sum of unknowns, we apply the same method to general scale invariant divergences . The multiplicative and non-multiplicative forms of these algorithms are shown in each case.\\
This allows us to obtain the algorithms corresponding to the divergences developed in sections 4 to 9 as specific applications of the general case.\\
The algorithms corresponding to the invariant divergences based on Tsallis entropy are detailed for comparison.\\
In the appendix we summarise the opposite of the gradients of the divergences corresponding to the different forms of entropy, in order to clearly highlight the analogies between them.

\section{Recalls (Very brief) on inverse problems.}
We dispose of a measurement of a physical quantity noted ``$y$'' which will be noted ``$p$'' and of a model describing this quantity; in our work we will consider that this model is linear ``$Hx$'', it will be noted ``$q$''.\\
Here, ``$H$'' is a known matrix and ``$x$'' is the true unknown of the problem.\\
It is obvious that in this context, ``$p$'' and ``$q$'' are not probability density functions, unlike the analogous quantities that appear in the field of Statistical Physics.\\
This particular point is one of the problems that arise when constructing divergences.\\
The unknown parameters ``$x$'' are subject to constraints of 2 types.\\
1 - the components of "$x$'' must be positive or zero.\\
2 - in some cases an additional constraint will be of the type $\sum_{i}x_{i}=C$, ``$C$'' being a known value.\\
Solving the problem implies the minimisation with respect to ``$x$'', under constraints, of a discrepancy function or ``\textit{divergence}'' between the measurement and the model ; such a divergence is generally noted as $D(p\|q)$.\\
This function is not necessarily symmetrical $D(p\|q)\neq D(q\|p)$, and in all generality, the important point is the convexity with respect to ``$q$'' which entails the convexity with respect to ``$x$''.\\
The divergence $D(q\|p)$ is called the dual divergence of $D(p\|q)$.\\
The construction of the different types of divergence is founded on the knowledge of a strictly convex function $f_{c}(x)$ (\textit{standard convex function}) with the following properties;
$f_{c}(1)=0$ and $f'_{c}(1)=0$.\\
Starting from a convex function $f(x)$, we obtain the ``\textit{standard convex function}'' $f_{c}(x)$, by the relation:
\begin{equation}
	f_{c}(x)=f(x)-f(1)-(x-1)f'(1)
\end{equation}
The divergences considered in this work are the Csiszär divergences and the Bregman divergences; the Jensen divergences proposed in \cite{lanteri2019}, \cite{lanteri2020}, will not be treated here.\\
Recall that the Csiszär divergence constructed on the strictly convex function $f(x)$ is expressed as:
\begin{equation}
	C(p\|q)=\sum_{i}q_{i}f\left(\frac{p_{i}}{q_{i}}\right)
	\label{dcsiszar}
\end{equation}
Similarly, the Bregman divergence is expressed as:
\begin{equation}
	B(p\|q)=\sum_{i}\left[f(p_{i})-f(q_{i})-(p_{i}-q_{i})f'(q_{i})\right]
	\label{dbregman}
\end{equation}
It should be noted that these divergences are separable (Trace form divergences).\\
For our problem, Csiszär divergences must be constructed on a standard convex function, they are jointly convex with respect to both arguments ($p$ and $q$).\\
On the other hand, Bregman divergences constructed on a simple convex function or on a standard convex function lead to the same divergence, they are convex with respect to the first argument, but may not be convex with respect to the second argument, which justifies an analysis of dual divergences.\\
Concerning the algorithmic aspect, we will rely on the SGM method exposed in \cite{lanteri2019}, \cite{lanteri2020}; this method allows us to easily take into account the non-negativity constraint, on the other hand, the sum constraint can be taken into account very simply with the same method, provided that we use divergences possessing the invariance property by change of scale on the variable ``$q$'', i.e. on the variable ``$x$'', given the linear model. \\
For this purpose, we will develop the ``$invariant$'' forms of the proposed divergences.

\section{Recalls (very brief) on entropy.}

In this note, we are interested in the different expressions of entropy and deformed logarithm that appear in the Statistical Physics literature \cite{tsallis2019}; the specific properties of entropies are not considered in this work.\\
In this context, by analogy with Shannon's entropy, we consider that the general expression of entropy is in the separable form (Trace form entropy):
\begin{equation}
	S=-\sum_{i}p_{i}\Lambda\left(p_{i}\right)=-\sum_{i}f(p_{i})
	\label{entropie classique}
\end{equation}
By definition, the ``$p_{i}$'' are probability densities, so they are in the range $0$ to $1$ and we have $\sum_{i}p_{i}=1$.\\
The function $\Lambda\left(p_{i}\right)$ designates the deformed Logarithm or the traditional Logarithm for the Shannon entropy.\\
Moreover, the entropy being a concave function, the functions $f(p_{i})$ are strictly convex with respect to ``$p_{i}$''.\\
Furthermore, the expressions of entropy appearing in the field of Statistical Physics depend on one or more parameters whose value domains are fixed so that the classical properties of entropy are respected (specifically in our case, the concavity).\\
From these expressions of entropy, using the definition (\ref{entropie classique}), we deduce the expressions of the corresponding deformed Logarithms $Log_{d}$.\\
In the field of Statistical Physics, relative entropy functions appear between 2 densities of probabilities, whereas in our case, we must define divergence functions analogous to relative entropies, but which relate to data fields which are not densities of probabilities, this implies some modifications in the way these functions are constructed. \\
For our particular purpose, we use the expressions of the various functions $f(x)$ defined in (\ref{entropie classique}) for $x\geq 0$ and not for $0\leq x\leq 1$.\\
The first question is: do the corresponding functions (specific to entropies) remain strictly convex over the whole extended domain?\\
The second question is: do they allow the construction of discrepancy functions between data fields that are not probability densities?\\  
The answer to these two questions is essential, because it is precisely on these two points that we will rely in order to build usable divergences here.\\
The third question is: are the expressions of $Log_{d}(x)$ deduced from the different entropies defined for all $x>0$ with the same parameters as the associated entropies, more precisely, are the basic properties of the classical logarithm maintained (concavity, behaviour in the neighbourhood of $0$, etc)?\\
This aspect is treated in detail in \cite{naudts2002}.\\
This last point is more anecdotic, because the properties of the deformed logarithm functions do not appear directly in our work, only the expressions of the entropies and the corresponding $f(x)$ functions are important.\\
In the following sections, we review the expressions of the entropies appearing in the field of Statistical Physics by taking first the case of the Shannon entropy which will be used as reference.

\section{Shannon entropy - Natural logarithm.}
The classical expression of the Shannon entropy related to the Boltzman-Gibbs statistic is expressed as:
\begin{equation}
	S(p)=-\sum_{i}p_{i}\log p_{i}
	\label{entropieshannon}
\end{equation}
Here, the logarithm that appears is the natural logarithm.\\
The strictly convex function that is involved is written:
\begin{equation}
	f(x)=x\log x
\end{equation}
This is not a standard convex function; the standard convex function deduced from $f(x)$ is written:
\begin{equation}
	f_{c}(x)=x\log x+1-x
\end{equation}

\subsection{Associated Csiszär divergence - Kullback-Leibler.}
On the basis of the function $f_{c}(x)$ defined in the previous section, we obtain by application of (\ref{dcsiszar}), the divergence:
\begin{equation}
	C(p\|q)=KL(p\|q)=\sum_{i}\left[p_{i}\log\frac{p_{i}}{q_{i}}-p_{i}+q_{i}\right]
	\label{divkl}
\end{equation}
This is the Kullback-Leibler divergence.\\
The opposite of its gradient with respect to ``$q$'' is written $\forall j$:
\begin{equation}
	-\frac{\partial KL(p\|q)}{\partial q_{j}}=\frac{p_{j}}{q_{j}}-1
	\label{mgradkl}
\end{equation}
This particular expression will be recovered for the various divergences examined here and will make it possible to highlight the modifications linked to the use of the ``\textsl{deformed logarithms}'' instead of  the natural logarithm.

\subsection{Dual Csiszär divergence.}
It can be constructed from (\ref{divkl}) by swapping the roles of ``$p$'' and ``$q$'' or by constructing a Csiszär divergence on the mirrored strictly convex function $\hat{f}_{c}(x)=xf_{c}\left(\frac{1}{x}\right)$.\\
It is written as:
\begin{equation}
	C(q\|p)=KL(q\|p)=\sum_{i}\left[q_{i}\log\frac{q_{i}}{p_{i}}-q_{i}+p_{i}\right]
	\label{divklduale}
\end{equation}
The opposite of its gradient with respect to ``$q$'' is written $\forall j$:
\begin{equation}
	-\frac{\partial KL(q\|p)}{\partial q_{j}}=\log\frac{p_{j}}{q_{j}}
\end{equation}
Note that this expression cannot be uniquely decomposed into a difference of 2 terms of the same sign because this decomposition changes depending on whether ``$p$'' and ``$q$'' are between ``$0$'' and ``$1$'', or greater than ``$1$''.

\subsection{Associated Bregman divergences.}
The Bregman divergence built on ``$f_{c}(x)$'' (or on ``$f(x)$'') leads from (\ref{dbregman}) to the Kullback-Leibler divergence given in the previous section; the same is true for dual divergences. This is the (only) common point between the Csiszär divergences and the Bregman divergences.

\section{Tsallis entropy \cite{tsallis1988}.}

With the change of notation ($q\leftrightarrow t$), the corresponding entropy will be written:
\begin{equation}
	S_{T}(p)=-\sum_{i}\frac{p^{t}_{i}-p_{i}}{t-1}\ \ \ ;\ \ \ t>0
	\label{entropietsallis}
\end{equation}
We will observe that by making the passage to the limit $t\rightarrow 1$ we recover the Shannon entropy.\\
With the definition (\ref{entropie classique}), the Tsallis deformed logarithm is written:
\begin{equation}
	\Lambda(x)=\log_{T}(x)=\frac{x^{t-1}-1}{t-1}
	\label{logtsallis}
\end{equation}
This function is concave for $t<2$; it is of course null for $x=1$.\\
However, one finds in the literature \cite{asgarani2015} another expression of the deformed Logarithm of Tsallis which is written:
\begin{equation}
	\log_{TT}(x)=\frac{x^{1-t}-1}{1-t}
\end{equation}
This function is equivalent to a logarithm for $t>0$.\\
If we associate to this form another expression of the entropy which is written:
\begin{equation}
	S(p)=-\sum_{i}p^{t}_{i}\log_{TT}(p_{i})
\end{equation}
We recover the entropy noted $S_{T}$ (\ref{entropietsallis}).\\
In what follows, we will use the form (\ref{logtsallis}) for the deformed logarithm of Tsallis and we will associate it with the classical form of the entropy (\ref{entropie classique}).\\
Anyway, the strictly convex function allowing to build divergences will be written:
\begin{equation}
	f(x)=\frac{x^{t}-x}{t-1}
\end{equation}
For this function, we have $f(1)=0$ but on the other hand $f'(1)=1$, it is not a standard convex function, it is consequently inadequate to construct Csiszär divergences useful in our problem.\\
The standard strictly convex function constructed on $f(x)$ is written:
\begin{equation}
	f_{c}(x)=\frac{x^{t}-x}{t-1}-x+1
\end{equation}

\subsection{Csiszär divergence founded on Tsallis entropy.}
The Csiszär divergence built on the strictly convex function $f_{c}(x)$ defined in the previous section is written, after all the simplifications:
\begin{equation}
	C_{T}(p\|q)=\frac{1}{t-1}\left[\sum_{i}p^{t}_{i}q^{1-t}_{i}-t\sum_{i}p_{i}+(t-1)\sum_{i}q_{i}\right]
	\label{ctsa}
\end{equation}
The opposite of the gradient is written, all calculations being done, $\forall j$:
\begin{equation}
	-\frac{\partial C_{T}(p\|q)}{\partial q_{j}}=\left(\frac{p_{j}}{q_{j}}\right)^{t}-1
\end{equation}
This can also be written:
\begin{equation}
	-\frac{\partial C_{T}(p\|q)}{\partial q_{j}}=\frac{p_{j}}{q_{j}}\left(\frac{p_{j}}{q_{j}}\right)^{t-1}-1
\end{equation}
With this expression, by making $t=1$, we immediately find the expression of the opposite of the gradient corresponding to the Kullback-Leibler divergence (\ref{mgradkl}) founded on the Shannon entropy.

\subsection{Dual Csiszär divergence founded on Tsallis entropy.}
It is immediately deduced from the expression (\ref{ctsa}) by performing the permutation $p\leftrightarrow q$, but it can be constructed directly in the sense of Csiszâr by relying on the standard strictly convex function $\hat{f}_{c}(x)$ (mirror function), deduced from $f_{c}(x)$ by the transformation:
\begin{equation}
	\hat{f}_{c}(x)=x f_{c}\left(\frac{1}{x}\right)
\end{equation}
It is written:
\begin{equation}
	C_{T}(q\|p)=\frac{1}{t-1}\left[\sum_{i}q^{t}_{i}p^{1-t}_{i}-t\sum_{i}q_{i}+(t-1)\sum_{i}p_{i}\right]
	\label{ctsaduale}                               
\end{equation}
The opposite of its gradient with respect to ``$q$'' is given $\forall j$ by:
\begin{equation}
	-\frac{\partial C_{T}(q\|p)}{\partial q_{j}}=\frac{t}{1-t}\left[\left(\frac{p_{j}}{q_{j}}\right)^{1-t}-1\right]
\end{equation}
Which can also be written:
\begin{equation}
	-\frac{\partial C_{T}(q\|p)}{\partial q_{j}}=\frac{1}{1-t}\left[t\left(\frac{p_{j}}{q_{j}}\right)^{1-t}-1\right]+1
\end{equation}
Passing to the limit $t\rightarrow 1$ allows to find the result obtained for the dual Csiszär divergence corresponding to the Shannon entropy.

\subsection{Bregman divergence founded on Tsallis entropy.}
The corresponding Bregman divergence can be constructed indifferently on the strictly convex function $f(x)$ or on the function $f_{c}(x)$; it is written:
\begin{equation}
B_{T}(p\|q)=\sum_{i}\left[q^{t}_{i}-\frac{1}{1-t}p^{t}_{i}+\frac{t}{1-t}p_{i}q^{t-1}_{i}\right]
\label{btsa}	
\end{equation}
The opposite of the gradient with respect to ``$q$'' is written all calculations performed $\forall j$:
\begin{equation}
-\frac{\partial B_{T}(p\|q)}{\partial q_{j}}=t\;q^{t-1}_{j}\left(\frac{p_{j}}{q_{j}}-1\right)	
\end{equation}

\subsection{Dual Bregman divergence founded on Tsallis entropy.}
It is immediately deduced from the expression (\ref{btsa}) by performing the permutation $p\leftrightarrow q$; it is expressed as:
\begin{equation}
B_{T}(q\|p)=\sum_{i}\left[p^{t}_{i}-\frac{1}{1-t}q^{t}_{i}+\frac{t}{1-t}q_{i}p^{t-1}_{i}\right]
\label{btsaduale}	
\end{equation}
The opposite of the gradient with respect to ``$q$'' is written all calculations completed $\forall j$:
\begin{equation}
-\frac{\partial B_{T}(q\|p)}{\partial q_{j}}=\frac{t}{1-t}\;q^{t-1}_{j}\left[\left(\frac{p_{j}}{q_{j}}\right)^{t-1}-1\right]
\end{equation}

\section{Kaniadakis entropy \cite{kaniadakis2001}, \cite{kaniadakis2002}, ``\textit{K}'' entropy.}

The expression of the Entropy is written:
\begin{equation}
	S_{K}(p)=-\sum_{i}p_{i}\log_{K}(p_{i})=-\sum_{i}\frac{p^{1+K}_{i}-p^{1-K}_{i}}{2K}\ \ \ ;\ \ \ -1<K<1
\end{equation}
It is invariant in the transformation $K\leftrightarrow -K$, which makes it possible to limit the domain of variations of $K$ to the interval $0<K<1$.\\
The Shannon entropy is obtained by making the passage to the limit $K\rightarrow 0$.\\
The strictly convex function deduced from this expression will be:
\begin{equation}
	f(x)=\frac{x^{1+K}-x^{1-K}}{2K}
\end{equation}
For this function, we have: $f(1)=0$, but $f'(1)=1$; it does not allow to construct in the sense of Csiszär, divergences usables in our context.\\
The standard convex function deduced from this one, allowing to build divergences adapted to our problems will be written:
\begin{equation}
	f_{c}(x)=\frac{x^{1+K}-x^{1-K}}{2K}-x+1
\end{equation}
The expression of the deformed logarithm of Kaniadakis is written:
\begin{equation}
	\log_{K}(x)=\frac{x^{K}-x^{-K}}{2K}
\end{equation}
This function is zero for $x=1$; it is increasing and concave for $-1<K<1$.

\subsection{Csiszär divergence built on the ``$K$'' entropy.}
The Csiszär divergence obtained from the standard convex function $f_{c}(x)$ is written, after simplification:
\begin{equation}
	C_{K}(p\|q)=\frac{1}{2K}\left[\sum_{i}p^{1+K}_{i}q^{-K}_{i}-\sum_{i}p^{1-K}_{i}q^{K}_{i}\right]-\sum_{i}p_{i}+\sum_{i}q_{i}
	\label{cska}
\end{equation}
The opposite of the gradient with respect to ``$q$'' is written $\forall j$:
\begin{equation}
	-\frac{\partial C_{K}(p\|q)}{\partial q_{j}}=\frac{1}{2}\left[\left(\frac{p_{j}}{q_{j}}\right)^{1+K}+\left(\frac{p_{j}}{q_{j}}\right)^{1-K}\right]-1
	\label{gradcska}
\end{equation}
This can still be written:
\begin{equation}
	-\frac{\partial C_{K}(p\|q)}{\partial q_{j}}=\frac{p_{j}}{q_{j}}\left[\frac{1}{2}\left(\frac{p_{j}}{q_{j}}\right)^{K}+\frac{1}{2}\left(\frac{p_{j}}{q_{j}}\right)^{-K}\right]-1
\end{equation}
With this writing, by making $K=0$, we immediately find the expression of the opposite of the gradient corresponding to the Kullback-Leibler divergence based on the Shannon entropy in which the deformed Logarithm becomes the classical Logarithm.

\subsection{Dual Csiszär divergence built on ``$K$'' entropy.}
It is deduced from the relation (\ref{cska}) by making the permutation $p\leftrightarrow q$; it is written:
\begin{equation}
	C_{K}(q\|p)=\frac{1}{2K}\left[\sum_{i}q^{1+K}_{i}p^{-K}_{i}-\sum_{i}q^{1-K}_{i}p^{K}_{i}\right]-\sum_{i}q_{i}+\sum_{i}p_{i}
	\label{ckad}
\end{equation}
The opposite of the gradient with respect to ``$q$'' is written $\forall j$:
\begin{equation}
	-\frac{\partial C_{K}(q\|p)}{\partial q_{j}}=\left[\frac{1-K}{2K}\left(\frac{p_{j}}{q_{j}}\right)^{K}-\frac{1+K}{2K}\left(\frac{p_{j}}{q_{j}}\right)^{-K}\right]+1
\end{equation}

\subsection{Bregman divergence founded on ``$K$'' entropy.}
The corresponding Bregman divergence can be constructed on the strictly convex function $f(x)$ or on the function $f_{c}(x)$; it is written:
\begin{align}
B_{K}(p\|q)=&\frac{1}{2K}\sum_{i}\left[p^{1+K}_{i}-p^{1-K}_{i}-q^{1+K}_{i}+q^{1-K}_{i}\right. \nonumber \\  & \left. -(1+K)p_{i}q^{K}_{i}+(1-K)p_{i}q^{-K}_{i}\right. \nonumber \\  & \left. +(1+K)q^{1+K}_{i}-(1-K)q^{1-K}_{i}\right]
\label{brka}	
\end{align}
The opposite of the gradient with respect to ``$q$'' is written all calculations performed $\forall j$:
\begin{equation}
-\frac{\partial B_{K}(p\|q)}{\partial q_{j}}=\left(\frac{p_{j}}{q_{j}}-1\right)\left[\frac{1+K}{2}q^{K}_{J}+\frac{1-K}{2}q^{-K}_{J}\right]	
\end{equation}

\subsection{Dual Bregman divergence built on ``$K$'' entropy.}
It is deduced from the relation (\ref{brka}) by performing the permutation $p\leftrightarrow q$; we obtain:
\begin{align}
B_{K}(q\|p)=&\frac{1}{2K}\sum_{i}\left[q^{1+K}_{i}-q^{1-K}_{i}-p^{1+K}_{i}+p^{1-K}_{i}\right. \nonumber \\  & \left. -(1+K)q_{i}p^{K}_{i}+(1-K)q_{i}p^{-K}_{i}\right. \nonumber \\  & \left. +(1+K)p^{1+K}_{i}-(1-K)p^{1-K}_{i}\right]
\label{brkaduale}	
\end{align}
The opposite of the gradient with respect to ``$q$'' is written all calculations performed $\forall j$:
\begin{equation}
-\frac{\partial B_{K}(q\|p)}{\partial q_{j}}=\left[\frac{1+K}{2K}\left(p_{j}^{K}-q^{K}_{j}\right)-\frac{1-K}{2K}\left(p_{j}^{-K}-q^{-K}_{j}\right)\right]
\end{equation}

\section{Abe's entropy \cite{abe1997}.}

This entropy is expressed as follows:
\begin{equation}
	S_{A}\left( p \right)=-\sum_{i}p_{i}\log_{A}\left(p_{i}\right)=-\sum_{i}\frac{p^{z}_{i}-p^{\frac{1}{z}}_{i}}{z-\frac{1}{z}}
	\label{entropieabe}
\end{equation}
It is invariant under the transformation $z\leftrightarrow z^{-1}$ and thus the domain of values of the parameter ``$z$'' can be restricted to the interval $0<z\leq 1$.\\
The Shannon entropy is recovered by making the passage to the limit $z\rightarrow 1$.\\
The strictly convex function allowing to build divergences will be written:
\begin{equation}
	f(x)=\frac{x^{z}-x^{\frac{1}{z}}}{z-\frac{1}{z}}
\end{equation}
For this function, we have $f(1)=0$ but on the other hand $f'(1)=1$, it is not a standard convex function, it is consequently inadequate to construct Csiszär divergences useful in our problem.\\
The standard strictly convex function constructed on $f(x)$ is written:
\begin{equation}
	f_{c}(x)=\frac{x^{z}-x^{\frac{1}{z}}}{z-\frac{1}{z}}-x+1
\end{equation}
With the definition (\ref{entropie classique}) the expression of the Abe's deformed Logarithm is written:
\begin{equation}
	\log_{A}(x)=\frac{x^{z-1}-x^{\frac{1}{z}-1}}{z-\frac{1}{z}}
	\label{logabe}
\end{equation}
This function is increasing and concave for $0.5<z<2$; it is invariant under the transformation $z\leftrightarrow z^{-1}$.\\

\subsection{Csiszär divergence founded on Abe entropy.}
The standard strictly convex function $f_{c}(x)$ defined in the previous section allows us to construct a Csiszär divergence adapted to our problem; it is written:
\begin{equation}
	C_{A}\left(p\|q\right)=\frac{z}{z^{2}-1}\left[\sum_{i}p^{z}_{i}q^{1-z}_{i}-\sum_{i}p^{\frac{1}{z}}_{i}q^{1-\frac{1}{z}}_{i}\right]-\sum_{i}p_{i}+\sum_{i}q_{i}
	\label{csabe}
\end{equation}
The opposite of the gradient with respect to ``$q$'' is written $\forall j$:
\begin{equation}
	-\frac{\partial C_{A}(p\|q)}{\partial q_{j}}=\frac{z}{z+1}\left(\frac{p_{j}}{q_{j}}\right)^{z}+\frac{1}{z+1}\left(\frac{p_{j}}{q_{j}}\right)^{\frac{1}{z}}-1
\end{equation}
Which can also be written:
\begin{equation}
	-\frac{\partial C_{A}(p\|q}{\partial q_{j}}=\frac{p_{j}}{q_{j}}\left[\frac{z}{z+1}\left(\frac{p_{j}}{q_{j}}\right)^{z-1}+\frac{1}{z+1}\left(\frac{p_{j}}{q_{j}}\right)^{\frac{1}{z}-1}\right]-1
\end{equation}
With this writing, by making $z=1$, we immediately recover the expression of the opposite of the gradient corresponding to the Kullback-Leibler divergence based on the Shannon entropy.

\subsection{Dual Csiszär divergence according to Abe entropy.}
It is deduced from the expression (\ref{csabe}) by performing the permutation $p\leftrightarrow q$; thus we obtain:
\begin{equation}
	C_{A}\left(q\|p\right)=\frac{z}{z^{2}-1}\left[\sum_{i}q^{z}_{i}p^{1-z}_{i}-\sum_{i}q^{\frac{1}{z}}_{i}p^{1-\frac{1}{z}}_{i}\right]-\sum_{i}q_{i}+\sum_{i}p_{i}
	\label{csabeduale}
\end{equation}
The opposite of the gradient with respect to ``$q$'' is written $\forall j$:
\begin{equation}
	-\frac{\partial C_{A}(q\|p)}{\partial q_{j}}=\frac{z}{1-z^{2}}\left[z\left(\frac{p_{j}}{q_{j}}\right)^{1-z}-\frac{1}{z}\left(\frac{p_{j}}{q_{j}}\right)^{1-\frac{1}{z}}\right]+1
\end{equation}
Which can also be written:
\begin{equation}
	-\frac{\partial C_{A}(q\|p)}{\partial q_{j}}=\frac{p_{j}}{q_{j}}\left[\frac{z^{2}}{1-z^{2}}\left(\frac{p_{j}}{q_{j}}\right)^{-z}-\frac{1}{1-z^{2}}\left(\frac{p_{j}}{q_{j}}\right)^{-\frac{1}{z}}\right]+1
\end{equation}

\subsection{Bregman divergence founded on Abe entropy.}
The corresponding Bregman divergence can be constructed on the strictly convex function $f(x)$ or on the function $f_{c}(x)$; it is written:
\begin{align}
B_{A}(p\|q)=&\frac{1}{z-\frac{1}{z}}\sum_{i}\left[p^{z}_{i}-p^{\frac{1}{z}}_{i}-q^{z}_{i}+q^{\frac{1}{z}}_{i}\right. \nonumber \\  & \left. -z p_{i}q^{(z-1)}_{i}+\frac{1}{z}p_{i}q^{(\frac{1}{z}-1)}_{i}\right. \nonumber \\  & \left. +zq^{z}_{i}-\frac{1}{z}q^{\frac{1}{z}}_{i}\right]
\label{brabe}	
\end{align}
The opposite of the gradient with respect to ``$q$'' is written $\forall j$:
\begin{equation}
-\frac{\partial B_{A}(p\|q)}{\partial q_{j}}=\left(\frac{p_{j}}{q_{j}}-1\right)\left[\frac{z^{2}}{z+1}q^{(z-1)}_{j}+\frac{1}{z(z+1)}q^{(\frac{1}{z}-1)}_{j}\right]	
\end{equation}

\subsection{Dual Bregman divergence founded on Abe entropy.}
It is deduced from the expression (\ref{brabe}) by permutation of ``$p$'' and ``$q$''; it is written:
\begin{align}
B_{A}(q\|p)=&\frac{1}{z-\frac{1}{z}}\sum_{i}\left[q^{z}_{i}-q^{\frac{1}{z}}_{i}-p^{z}_{i}+p^{\frac{1}{z}}_{i}\right. \nonumber \\  & \left. -z q_{i}p^{(z-1)}_{i}+\frac{1}{z}q_{i}p^{(\frac{1}{z}-1)}_{i}\right. \nonumber \\  & \left. +zp^{z}_{i}-\frac{1}{z}p^{\frac{1}{z}}_{i}\right]
\label{brabeduale}	
\end{align}
The opposite of the gradient with respect to ``$q$'' is written $\forall j$:
\begin{equation}
-\frac{\partial B_{A}(q\|p)}{\partial q_{j}}=\left[\frac{z^{2}}{z^{2}-1}(p^{z-1}_{j}-q^{z-1}_{j})-\frac{1}{z^{2}-1}(p^{\frac{1}{z}-1}_{j}-q^{\frac{1}{z}-1}_{j})\right]	
\end{equation}

\section{Gamma \textbf{($\gamma$)} entropy.}
 
The ``$\gamma$'' entropy mentioned in  \cite{kaniadakis2005}, is written:
\begin{equation}
	S_{\gamma}(p)=-\sum_{i}p_{i}\log_{\gamma}(p_{i})=-\sum_{i}\frac{p^{1+2\gamma}_{i}-p^{1-\gamma}_{i}}{3\gamma}
\end{equation}
This function is defined for $-0.5\leq \gamma\leq 1$.\\
The passage to the limit $\gamma\rightarrow 0$ allows to recover the entropy of Shannon.\\
The strictly convex function allowing to build divergences is written:
\begin{equation}
	f(x)=\frac{x^{1+2\gamma}-x^{1-\gamma}}{3\gamma}
\end{equation}                            
For this function, we have $f(1)=0$ but on the other hand $f'(1)=1$, it is not a standard convex function, it is consequently inadequate to construct Csiszär divergences useful in our problem.\\
The standard strictly convex function constructed on $f(x)$ is written:
\begin{equation}
	f_{c}(x)=\frac{x^{1+2\gamma}-x^{1-\gamma}}{3\gamma}+1-x
\end{equation}                            
With the definition (\ref{entropie classique}) the expression of the ``$\gamma$'' deformed logarithm is written:
\begin{equation}
	\log_{\gamma}(x)=\frac{x^{2\gamma}-x^{-\gamma}}{3\gamma}
\end{equation}
As stated in Kaniadakis et al. \cite{kaniadakis2005}, this function is zero for $x=1$, it is increasing and concave for $-0.5<\gamma <0.5$.\\

\subsection{Csiszär divergence built on the "$\gamma$" entropy .}
The function $f_{c}(x)$ defined in the previous section allows us to construct the following Csiszär divergence:
\begin{equation}
	C_{\gamma}(p\|q)=\frac{1}{3\gamma}\left[\sum_{i}p^{1+2\gamma}_{i}q^{-2\gamma}_{i}-\sum_{i}p^{1-\gamma}_{i}q^{\gamma}_{i}\right]-\sum_{i}p_{i}+\sum_{i}q_{i}
	\label{csgam}
\end{equation}
The opposite of the gradient with respect to ``$q$'' is written $\forall j$:
\begin{equation}
	-\frac{\partial C_{\gamma}(p\|q)}{\partial q_{j}}=\frac{2}{3}\left(\frac{p_{j}}{q_{j}}\right)^{1+2\gamma}+\frac{1}{3}\left(\frac{p_{j}}{q_{j}}\right)^{1-\gamma}-1
\end{equation}
Which can also be written:
\begin{equation}
	-\frac{\partial C_{\gamma}(p\|q)}{\partial q_{j}}=\frac{p_{j}}{q_{j}}\left[\frac{2}{3}\left(\frac{p_{j}}{q_{j}}\right)^{2\gamma}+\frac{1}{3}\left(\frac{p_{j}}{q_{j}}\right)^{-\gamma}\right]-1
\end{equation}
With this writing, by setting $\gamma=0$, we immediately recover the expression of the opposite of the gradient corresponding to the Kullback-Leibler divergence based on the Shannon entropy.

\subsection{Dual Csiszär divergence based on the ``$\gamma$'' entropy.}
It is constructed from the expression (\ref{csgam}) by swapping ``$p$'' and ``$q$''; it is written:
\begin{equation}
	C_{\gamma}(q\|p)=\frac{1}{3\gamma}\left[\sum_{i}q^{1+2\gamma}_{i}p^{-2\gamma}_{i}-\sum_{i}q^{1-\gamma}_{i}p^{\gamma}_{i}\right]-\sum_{i}q_{i}+\sum_{i}p_{i}
	\label{csgamduale}
\end{equation}
The opposite of the gradient with respect to ``$q$'' is written $\forall j$:
\begin{equation}
	-\frac{\partial C_{\gamma}(q\|p)}{\partial q_{j}}=\frac{1}{3\gamma}\left[(1-\gamma)\left(\frac{p_{j}}{q_{j}}\right)^{\gamma}-(1+2\gamma)\left(\frac{p_{j}}{q_{j}}\right)^{-2\gamma}\right]+1
\end{equation}
Or also:
\begin{equation}
	-\frac{\partial C_{\gamma}(q\|p)}{\partial q_{j}}=\frac{p_{j}}{q_{j}}\left[\frac{1-\gamma}{3\gamma}\left(\frac{p_{j}}{q_{j}}\right)^{\gamma-1}-\frac{1+2\gamma}{3\gamma}\left(\frac{p_{j}}{q_{j}}\right)^{-2\gamma-1}\right]+1
\end{equation}

\subsection{Bregman divergence built on the ``$\gamma$'' entropy.}
The corresponding Bregman divergence can be constructed indifferently on the strictly convex functions $f(x)$ or $f_{c}(x)$; it is written:
\begin{align}
B_{\gamma}(p\|q)=&\frac{1}{3 \gamma}\sum_{i}\left[p^{2 \gamma+1}_{i}-p^{1-\gamma}_{i}-q^{2 \gamma+1}_{i}+q^{1-\gamma}_{i}\right. \nonumber \\  & \left. -(2 \gamma+1) p_{i}q^{2 \gamma}_{i}+(1-\gamma)p_{i}q^{-\gamma}_{i}\right. \nonumber \\  & \left. +(2 \gamma+1)q^{2 \gamma+1}_{i}-(1-\gamma)q^{1-\gamma}_{i}\right]
\label{brgam}	
\end{align}
The opposite of the gradient with respect to ``$q$'' is written $\forall j$:
\begin{equation}
-\frac{\partial B_{\gamma}(p\|q)}{\partial q_{j}}=\left(\frac{p_{j}}{q_{j}}-1\right)\left[\frac{2(2\gamma+1)}{3}q^{2 \gamma}_{j}+\frac{1-\gamma}{3}q^{-\gamma}_{j}\right]	
\end{equation}

\subsection{Dual Bregman divergence based on the ``$\gamma$'' entropy.}
It is constructed from the relation (\ref{brgam}) by swapping ``$p$'' and ``$q$''; it is written:
\begin{align}
B_{\gamma}(q\|p)=&\frac{1}{3 \gamma}\sum_{i}\left[q^{2 \gamma+1}_{i}-q^{1-\gamma}_{i}-p^{2 \gamma+1}_{i}+p^{1-\gamma}_{i}\right. \nonumber \\  & \left. -(2 \gamma+1) q_{i}p^{2 \gamma}_{i}+(1-\gamma)q_{i}p^{-\gamma}_{i}\right. \nonumber \\  & \left. +(2 \gamma+1)p^{2 \gamma+1}_{i}-(1-\gamma)p^{1-\gamma}_{i}\right]
\label{brgamduale}	
\end{align}
The opposite of the gradient with respect to ``$q$'' is written $\forall j$:
\begin{equation}
-\frac{\partial B_{\gamma}(q\|p)}{\partial q_{j}}=\left[\frac{2\gamma+1}{3\gamma}\left(p^{2\gamma}_{j}-q^{2\gamma}_{j}\right)-\frac{1-\gamma}{3\gamma}\left(p^{-\gamma}_{j}-q^{-\gamma}_{j}\right)\right]
\end{equation}

\section{Two parameters entropy \cite{kaniadakis2005}, K.L.S.}
The expression of the 2-parameter entropy $(K,r)$ proposed by Kaniadakis et al. \cite{kaniadakis2005}, is written:
\begin{equation}
	S_{K,r}(p)=-\sum_{i}p_{i}\log_{(K,r)}(p_{i})=-\sum_{i}p^{1+r}_{i}\frac{p^{K}_{i}-p^{-K}_{i}}{2K}
\end{equation}
To recover Shannon's entropy, we proceed in 2 steps: first we make $r=0$, then we perform the passage to the limit $K\rightarrow 0$.\\
With the definition (\ref{entropie classique}), the expression of the 2-parameters deformed Logarithm $(K,r)$ proposed by Kaniadakis et al. \cite{kaniadakis2005} is written:
\begin{equation}
	Log_{(K,r)}(x)=x^{r}\frac{x^{K}-x^{-K}}{2K}
\end{equation}
This function is increasing if $-\left|K\right|\leq r\leq \left|K\right|$.\\
It is concave:\\
- for $-\left|K\right|\leq r\leq \left|K\right|$ if $\left|K\right|\leq \frac{1}{2}$\\
or else,\\
- for $-\left|K\right|\leq r\leq 1-\left|K\right|$ if $\left|K\right|\geq \frac{1}{2}$.\\

The basis convex function deduced from the expression of the entropy is written:
\begin{equation}
	f(x)=\frac{x^{1+K+r}-x^{1-K+r}}{2K}
\end{equation}
This is not a standard convex function because $f'(1)=1$; the standard convex function deduced from $f(x)$ will be written:
\begin{equation}
	f_{c}(x)=\frac{x^{1+K+r}-x^{1-K+r}}{2K}+1-x
\end{equation}
 
\subsection{Csiszär divergence derived from the 2-parameter entropy.}
The Csiszär divergence derived from the function $f_{c}(x)$ defined in the previous section is written:
\begin{equation}
	C_{KLS}(p\|q)=\frac{1}{2K}\left[\sum_{i}p^{1+r+K}_{i}q^{-r-K}_{i}-\sum_{i}p^{1+r-K}_{i}q^{-r+K}_{i}\right]-\sum_{i}p_{i}+\sum_{i}q_{i}
	\label{cskls}
\end{equation}
The opposite of the gradient with respect to ``$q$'' is written all calculations performed $\forall j$:
\begin{equation}
	-\frac{\partial C_{KLS}(p\|q)}{\partial q_{j}}=\frac{1}{2K}\left[(K+r)\left(\frac{p_{j}}{q_{j}}\right)^{1+r+K}+(K-r)\left(\frac{p_{j}}{q_{j}}\right)^{1+r-K}\right]-1
\end{equation}
Which can also be written:
\begin{equation}
	-\frac{\partial C_{KLS}(p\|q)}{\partial q_{j}}=\left(\frac{p_{j}}{q_{j}}\right)\left[\frac{K+r}{2K}\left(\frac{p_{j}}{q_{j}}\right)^{r+K}+\frac{K-r}{2K}\left(\frac{p_{j}}{q_{j}}\right)^{r-K}\right]-1
\end{equation}
If, from this expression, we make $r=0$, we recover what we obtained for the deformed Logarithm of Kaniadakis (\ref{gradcska}), then by making $K\rightarrow 0$ we recover the case corresponding to the classical Logarithm and to the divergence of KullbacK Leibler.\\
The interesting form of the algorithm will be obtained under the condition that $-K\leq r\leq K$.

\subsection{Dual Csiszär divergence based on 2-parameter entropy.}
It is derived from the expression (\ref{cskls}) by swapping "$p$" and "$q$"; it is written:
\begin{equation}
	C_{KLS}(q\|p)=\frac{1}{2K}\left[\sum_{i}q^{1+r+K}_{i}p^{-r-K}_{i}-\sum_{i}q^{1+r-K}_{i}p^{-r+K}_{i}\right]-\sum_{i}q_{i}+\sum_{i}p_{i}
	\label{csklsduale}
\end{equation}
The opposite of the gradient with respect to ``$q$'' is written all computations done $\forall j$:
\begin{equation}
	-\frac{\partial C_{KLS}(q\|p)}{\partial q_{j}}=\frac{1}{2K}\left[(1+r-K)\left(\frac{p_{j}}{q_{j}}\right)^{K-r}-(1+r+K)\left(\frac{p_{j}}{q_{j}}\right)^{-K-r}\right]+1
\end{equation}
Or also:
\begin{equation}
	-\frac{\partial C_{KLS}(q\|p)}{\partial q_{j}}=\frac{p_{j}}{q_{j}}\left[\frac{1+r-K}{2K}\left(\frac{p_{j}}{q_{j}}\right)^{K-r-1}-\frac{1+r+K}{2K}\left(\frac{p_{j}}{q_{j}}\right)^{-K-r-1}\right]+1
\end{equation}

\subsection{Bregman divergence derived from the 2-parameter entropy.}
The Bregman divergence built on the functions $f(x)$ or $f_{c}(x)$ defined in the previous section is written:
\begin{align}
	B_{KLS}(p\|q)=\frac{1}{2K}&\sum_{i}\left[p^{1+r+K}_{i}-p^{1+r-K}_{i}-q^{1+r+K}_{i}+q^{1+r-K}_{i}\right. \nonumber \\  & \left.-(1+r+K)p_{i}q^{r+K}_{i}+(1+r+K)q^{1+r+K}_{i}\right. \nonumber \\  & \left.+(1+r-K)p_{i}q^{r-K}_{i}-(1+r-K)q^{1+r-K}_{i}\right]
	\label{brkls}
\end{align}
The opposite of the gradient with respect to ``$q$'' is written after some calculations $\forall j$:
\begin{align}
	-\frac{\partial B_{KLS}(p\|q)}{\partial q_{j}}=\left(\frac{p_{j}}{q_{j}}-1\right)&\left[\frac{(1+r+K)(r+K)}{2K}q^{r+K}_{j}\right. \nonumber \\  & \left.-\frac{(1+r-K)(r-K)}{2K}q^{r-K}_{j}\right]
\end{align}
By setting: $a=1+r+K$ and $b=1+r-K$ we obtain a general divergence (\ref{brgen}) which will be analyzed later on.

\subsection{Dual Bregman divergence built on the 2-parameter entropy.}
It is derived from the relation (\ref{brkls}) by permuting "$p$" and "$q$",
\begin{align}
	B_{KLS}(q\|p)=\frac{1}{2K}&\sum_{i}\left[q^{1+r+K}_{i}-q^{1+r-K}_{i}-p^{1+r+K}_{i}+p^{1+r-K}_{i}\right. \nonumber \\  & \left.-(1+r+K)q_{i}p^{r+K}_{i}+(1+r+K)p^{1+r+K}_{i}\right. \nonumber \\  & \left.+(1+r-K)q_{i}p^{r-K}_{i}-(1+r-K)p^{1+r-K}_{i}\right]
	\label{brklsduale}
\end{align}
The opposite of the gradient with respect to ``$q$'' is written $\forall j$:
\begin{equation}
-\frac{\partial B_{KLS}(q\|p)}{\partial q_{j}}=\left(\frac{1+r+K}{2K}\right)\left(p^{r+K}_{j}-q^{r+K}_{j}\right)-\left(\frac{1+r-K}{2K}\right)\left(p^{r-K}_{j}-q^{r-K}_{j}\right)
\end{equation}

\section{Other definitions of entropy and deformed logarithm.}
                            
In this section, we consider other ways of defining entropy which allow us to retrieve the expressions of paragraphs 4 to 9 and to give a very general form of entropy which groups them together.\\
The expressions of entropy which will be mentioned in paragraph 11 do not enter in this context.\\  
In the previous sections the form of the entropy used was defined by the relation:
\begin{equation}
	S(p)=-\sum_{i}p_{i}\Lambda(p_{i})
\end{equation}
where $\Lambda(p_{i})$ stands for a form of the Generalized Logarithm; the classical Logarithm corresponding to the Shannon entropy.\\
Another definition of the Shannon entropy proposed by Abe \cite{abe1997} is as follows:
\begin{equation}
	S(p)=\left[-\frac{d}{d\alpha}\sum_{i}p^{\alpha}_{i}\right]_{\alpha=1}\equiv \left[-\frac{d}{d\alpha}f(\alpha)\right]_{\alpha=1}
	\label{defabe}
\end{equation}
In this expression the derivative involved is the usual derivative.\\
From (\ref{defabe}), it is proposed to deform the classical derivative and to replace it by the Jackson ``$q$'' differential \cite{jackson1909} \cite{jackson1910} which is written, by changing the notation ($q \leftrightarrow t$): 
 \begin{equation}
	\frac{df(\alpha)}{d(\alpha:t)}=\frac{f(\alpha t)-f(\alpha)}{\alpha t-\alpha} 
 \end{equation}
(One could almost say that the additive increase of the variable in the classical derivative has been replaced by a multiplicative increase of the variable.)\\
With this definition, we get:
\begin{equation}
	S(p)=\left[-\frac{\sum_{i}p^{\alpha t}_{i}-\sum_{i}p^{\alpha}_{i}}{\alpha t-\alpha}\right]_{\alpha=1}
\end{equation}
Which corresponds to Tsallis' entropy \cite{tsallis1988}, already mentioned in (\ref{entropietsallis}):
\begin{equation}
	S_{T}(p)=-\frac{\sum_{i}p^{t}_{i}-\sum_{i}p_{i}}{t-1}=-\frac{\sum_{i}p^{t}_{i}-1}{t-1}
\end{equation}
The corresponding deformed Logarithm would be written (\ref{logtsallis}):
\begin{equation}
	Log_{T}(x)=\frac{x^{t-1}-1}{t-1}
\end{equation}
An alternative consists, still from (\ref{defabe}), in adopting another definition of the differential given by Mc Anally \cite{mcanally1995} which is written:
\begin{equation}
	\frac{df(\alpha)}{d(\alpha:t)}=\frac{f(\alpha t)-f(\alpha t^{-1})}{\alpha t-\alpha t^{-1}}
\end{equation}
We thus have an invariance $t\leftrightarrow t^{-1}$ and we obtain:
\begin{equation}
	S(p)=\left[-\frac{\sum_{i}p^{\alpha t}_{i}-\sum_{i}p^{\alpha t^{-1}}_{i}}{\alpha t-\alpha t^{-1}}\right]_{\alpha=1}=-\frac{\sum_{i}p^{t}_{i}-\sum_{i}p^{t^{-1}}_{i}}{t- t^{-1}}
\end{equation}
We thus recover the entropy of Abe (\ref{entropieabe}); the corresponding deformed Logarithm is written (\ref{logabe}):
\begin{equation}
	Log_{A}(x)=\frac{x^{t-1}-x^{t^{-1}-1}}{t-t^{-1}}
\end{equation}
Another form of generalized 2-parameter differential, initially used by Chakrabarti and Jagannathan \cite{chakrabarti1991}, is proposed by Borges and Roditi \cite{borges1998}; it consists in using the following definition:
\begin{equation}
	\frac{d_{ab}f(\alpha)}{d_{ab}(\alpha)}=\frac{f(a \alpha)-f(b \alpha)}{a \alpha-b \alpha}
\end{equation}

This form extends the definitions previously proposed, it leads, with the definition (\ref{defabe}) to the expression of the entropy:
\begin{equation}
	S_{ab}(p)=\left[-\frac{\sum_{i}p^{a \alpha}_{i}-\sum_{i}p^{b \alpha}_{i}}{a \alpha-b \alpha}\right]_{\alpha=1}=-\frac{\sum_{i}p^{a}_{i}-\sum_{i}p^{b}_{i}}{a-b}
	\label{entgen}
\end{equation}
This very general form and the associated deformed logarithm, also mentionned by Wada and Scarfone \cite{wada2010}, was founded on previous works of Sharma and Taneja \cite{sharma1975} and Mittal \cite{mittal1975}; it allows of course to recover the previous expressions of paragraphs 4 to 9.\\
The domains of values of the parameters ``$a$'' and ``$b$'' indicated in \cite{furuichi2010} and \cite{borges1998} are the following:
\begin{equation}
	0\leq a\leq 1\leq b
	\label{valab}
\end{equation}
or also:
\begin{equation}
 0\leq b\leq 1\leq a
\label{valabbis}
\end{equation}
The expression of the deformed Logarithm deduced from this form of the entropy is written:
\begin{equation}
	Log_{ab}(x)=\frac{x^{a-1}-x^{b-1}}{a-b}
\end{equation}
The deformed logarithm thus defined has the concavity properties of the classical logarithm in a domain of parameter values given by:
\begin{equation}
	0\leq a\leq 1\leq b\leq 2\ \ \ \ ;\ \ \ \ 0\leq b\leq 1\leq a\leq 2
\end{equation}

\section{Divergences based on the general form of the entropy (\ref{entgen}).}
We develop in the following paragraphs the various forms of divergences based on the general entropy (\ref{entgen}).\\
The particular cases developed in sections 4 to 9 can be found by making the following adaptations.\\
\begin{itemize}
\item Shannon entropy: $a\rightarrow 1$ et  $b\rightarrow 1$ taking into account (\ref{valab}) or (\ref{valabbis})\\
\item Tsallis entropy: $a=t,$\ \ \ $  b=1$\\
\item Kaniadakis entropy: $a=1+K,$\ \ \ $ b=1-K$\\
\item Abe entropy: $a=z,$\ \ \ $ b=\frac{1}{z}$\\
\item ``$\gamma$'' entropy: $a=2\gamma+1,$\ \ \ $ b=1-\gamma$\\
\item 2 parameters (KLS) entropy: $a=1+r+K,$\ \ \ $ b=1+r-K$\\
\end{itemize}

\subsection{Csiszär divergence based on the general entropy.}

The strictly convex basis function deduced from the general form of entropy (\ref{entgen}) is written:
\begin{equation}
	f(x)=\frac{x^{a}-x^{b}}{a-b}
\end{equation}
This is not a standard function because $f'(1)=1$; the standard strictly convex function deduced from $f(x)$ is written:
\begin{equation}
	f_{c}(x)=\frac{x^{a}-x^{b}}{a-b}-x+1
\end{equation}
The Csiszär divergence based on this function is written:
\begin{equation}
	C_{ab}(p\|q)=\frac{1}{a-b}\left[\sum_{i}p^{a}_{i}q^{1-a}_{i}-\sum_{i}p^{b}_{i}q^{1-b}_{i}\right]-\sum_{i}p_{i}+\sum_{i}q_{i}
	\label{csgen}
\end{equation}
From this expression, using the inequalities (\ref{valab}) or (\ref{valabbis}), we recover the Kullback-Leibler divergence by making the passage to the limit $a\rightarrow 1$ and $b\rightarrow 1$.\\
The opposite of its gradient with respect to $q$ is written as $forall j$:
\begin{equation}
	-\frac{\partial C_{ab}(p\|q)}{\partial q_{j}}=\left[\frac{a-1}{a-b}\left(\frac{p_{j}}{q_{j}}\right)^{a}+\frac{1-b}{a-b}\left(\frac{p_{j}}{q_{j}}\right)^{b}\right]-1
	\label{mgradcsgen}
\end{equation}
What can still be written:
\begin{equation}
	-\frac{\partial C_{ab}(p\|q)}{\partial q_{j}}=\left(\frac{p_{j}}{q_{j}}\right)\left[\frac{a-1}{a-b}\left(\frac{p_{j}}{q_{j}}\right)^{a-1}+\frac{1-b}{a-b}\left(\frac{p_{j}}{q_{j}}\right)^{b-1}\right]-1
	\label{mgradcsgenbis}
\end{equation}
The decomposition of the opposite of the gradient allowing to obtain a multiplicative form of the inversion algorithm implies:
\begin{equation}
	\frac{a-1}{a-b}>0\ \ \ ;\ \ \ \frac{1-b}{a-b}>0
\end{equation}
These inequalities are fulfilled if ``$a$'' and ``$b$'' belong to the domains of values given by (\ref{valab}) or (\ref{valabbis}).

\subsection{Dual Csiszär divergence built on general entropy.}
It is derived from the expression (\ref{csgen}) by swapping ``$p$'' and ``$q$''; it is written:
\begin{equation}
	C_{ab}(q\|p)=\frac{1}{a-b}\left[\sum_{i}q^{a}_{i}p^{1-a}_{i}-\sum_{i}q^{b}_{i}p^{1-b}_{i}\right]-\sum_{i}q_{i}+\sum_{i}p_{i}
	\label{csgenduale}
\end{equation}
The opposite of its gradient with respect to ``$q$'' is written all calculations performed $\forall j$:
\begin{equation}
	-\frac{\partial C_{ab}(q\|p)}{\partial q_{j}}=\frac{1}{a-b}\left[b\left(\frac{p_{j}}{q_{j}}\right)^{1-b}-a\left(\frac{p_{j}}{q_{j}}\right)^{1-a}\right]+1
\end{equation}
or also:
\begin{equation}
	-\frac{\partial C_{ab}(q\|p)}{\partial q_{j}}=\frac{p_{j}}{q_{j}}\left[\frac{b}{a-b}\left(\frac{p_{j}}{q_{j}}\right)^{-b}-\frac{a}{a-b}\left(\frac{p_{j}}{q_{j}}\right)^{-a}\right]+1
\end{equation} 
\textbf{Warning: the decomposition of the opposite of the gradient into a difference of 2 positive terms changes according to the sign of $a-b$.}

\subsection{Bregman divergence based on the general entropy.}
This divergence can be constructed either on the basis of the function $f(x)$ or on the basis of $f_{c}(x)$; the expression obtained is identical and can be written with all calculations performed:
\begin{align}
B_{ab}(p\|q)=\frac{1}{a-b}\sum_{i}&\left[p^{a}_{i}-p^{b}_{i}+(a-1)q^{a}_{i}-(b-1)q^{b}_{i}\right. \nonumber \\  & \left.-ap_{i}q^{a-1}_{i}+bp_{i}q^{b-1}_{i}\right]
\label{brgen}	
\end{align}
The opposite of the gradient with respect to ``$q$'' is written, $\forall j$, after factorization:
\begin{equation}
	-\frac{\partial B_{ab}(p\|q)}{\partial q_{j}}=\left(\frac{p_{j}}{q_{j}}-1\right)\left[\frac{(a^{2}-a)}{a-b}q^{a-1}_{j}+\frac{(b-b^{2})}{a-b}q^{b-1}_{j}\right]
	\label{mgradbrgen}
\end{equation}
Considering the ranges of values of ``$a$'' and ``$b$'' defined in (\ref{valab}) or (\ref{valabbis}), we have:
\begin{equation}
	\frac{a^{2}-a}{a-b}>0\ \ \ ;\ \ \ \frac{b-b^{2}}{a-b}>0
\end{equation}
This will allow a purely multiplicative form of the inversion algorithm to be written without ambiguity.

\subsection{Dual Bregman divergence based on the general entropy.}
It is derived from the expression (\ref{brgen}) by swapping ``$p$'' and ``$q$''; it is written:
\begin{align}
B_{ab}(q\|p)=\frac{1}{a-b}\sum_{i}&\left[q^{a}_{i}-q^{b}_{i}+(a-1)p^{a}_{i}-(b-1)p^{b}_{i}\right. \nonumber \\  & \left.-aq_{i}p^{a-1}_{i}+bq_{i}p^{b-1}_{i}\right]
\label{brgenduale}	
\end{align}
The opposite of the gradient with respect to ``$q$'' is written, $\forall j$:
\begin{equation}
	-\frac{\partial B_{ab}(q\|p)}{\partial q_{j}}=\frac{1}{a-b}\left[a\left(p^{a-1}_{j}-q^{a-1}_{j}\right)-b\left(p^{b-1}_{j}-q^{b-1}_{j}\right)\right]
\end{equation}
Which can also be written:
\begin{equation}
	-\frac{\partial B_{ab}(q\|p)}{\partial q_{j}}=\frac{1}{a-b}\left[(a p^{a-1}_{j}+b q^{b-1}_{j})-(a q^{a-1}_{j}+b p^{b-1}_{j})\right]
\end{equation}
\textbf{Warning: the decomposition of the opposite of the gradient into a difference of 2 positive terms changes according to the sign of $a-b$.}

\section{Some other expressions of the deformed logarithm.}
There are some other expressions of the deformed logarithm founded in the literature that do not agree with the general form described in the previous sections. They are cited by Trivellato \cite{trivellato2013} and are considered here.
 
\subsection{Newton's deformed logarithm  \cite{newton2012}.}
It is given by the expression:
\begin{equation}
	L_{N}(x)=\frac{1}{2}\left(x-1+Log x\right)
\end{equation}
According to the relation (\ref{entropie classique}) we obtain for the corresponding entropy, the specific expression:
\begin{equation}
	S_{N}(p)=-\sum_{i}p_{i}L_{N}(p_{i})=-\sum_{i}\frac{1}{2}\left[p^{2}_{i}-p_{i}+p_{i}\:Log\: p_{i}\right]
\end{equation}
The basic strictly convex function is written accordingly:
\begin{equation}
	f(x)=\frac{1}{2}\left(x^{2}-x+x\:Log\: x\right)
\end{equation}
It is not a standard function, indeed $f^{'}(1)=1$.\\
The standard strictly convex function that we deduce from this is written:
\begin{equation}
	f_{c}(x)=\frac{1}{2}\left(x^{2}-3x+2+x\:Log\: x\right)
\end{equation}

\subsubsection{Csiszär divergence corresponding to Newton's entropy.}
The Csiszär divergence based on the previous $f_{c}(x)$ function is written:
\begin{equation}
	C_{N}(p\|q)=\frac{1}{2}\sum_{i}\left[p^{2}_{i}q^{-1}_{i}-3p_{i}+2q_{i}+p_{i}Log\left(\frac{p_{i}}{q_{i}}\right)\right]
	\label{csnew}
\end{equation}
The opposite of the gradient with respect to ``$q$'' is written $\forall j$:
\begin{equation}
	-\frac{\partial C_{N}(p\|q)}{\partial q_{j}}=\frac{p_{j}}{q_{j}}\left[\frac{1}{2}\frac{p_{j}}{q_{j}}+\frac{1}{2}\right]-1
\end{equation}
It is obviously equal to ``0'' if $p_{j}=q_{j}\ \ \forall j$.

\subsubsection{Dual Csiszär divergence corresponding to the Newton entropy.}
It is derived from the expression (\ref{csnew}) by swapping ``$p$'' and ``$q$''; it is written:
\begin{equation}
	C_{N}(q\|p)=\frac{1}{2}\sum_{i}\left[q^{2}_{i}p^{-1}_{i}-3q_{i}+2p_{i}+q_{i}Log\left(\frac{q_{i}}{p_{i}}\right)\right]
	\label{csnewduale}
\end{equation}
The opposite of the gradient with respect to ``$q$'' is written $\forall j$:
\begin{equation}
	-\frac{\partial C_{N}(q\|p)}{\partial q_{j}}=\frac{1}{2}\log\frac{p_{j}}{q_{j}}+1-\frac{q_{j}}{p_{j}}
\end{equation}
It is of course equal to ``0'' if $p_{j}=q_{j}\ \ \forall j$.

\subsubsection{Bregman divergence corresponding to Newton's entropy.}
The Bregman divergence can be obtained from the strictly convex function $f(x)$ or from the standard strictly convex function $f_{c}(x)$; we obtain:
\begin{equation}
B_{N}(p\|q)=\frac{1}{2}\sum_{i}\left[p^{2}_{i}-p_{i}+p_{i}\log p_{i}+q^{2}_{i}+q_{i}-2p_{i}q_{i}-p_{i}\log q_{i}\right]
\label{brnew}	
\end{equation}
The opposite of the gradient with respect to ``$q$'' is written $\forall j$:
\begin{equation}
	-\frac{\partial B_{N}(p\|q)}{\partial q_{j}}=\left(\frac{p_{j}}{q_{j}}-1\right)\left(\frac{1}{2}+q_{j}\right)
\end{equation}

\subsubsection{Dual Bregman divergence corresponding to Newton's entropy.}
It is deduced from the relation (\ref{brnew}) by swapping ``$p$'' and ``$q$''; it is written;
\begin{equation}
B_{N}(q\|p)=\frac{1}{2}\sum_{i}\left[q^{2}_{i}-q_{i}+q_{i}\log q_{i}+p^{2}_{i}+p_{i}-2q_{i}p_{i}-q_{i}\log p_{i}\right]
\label{brnewduale}	
\end{equation}
The opposite of the gradient with respect to ``$q$'' is written $\forall j$:
\begin{equation}
	-\frac{\partial B_{N}(q\|p)}{\partial q_{j}}=q_{j}\left(\frac{p_{j}}{q_{j}}-1\right)+\frac{1}{2}\log\frac{p_{j}}{q_{j}}
\end{equation}
This can also be written as:
\begin{equation}
	-\frac{\partial B_{N}(q\|p)}{\partial q_{j}}=\left(p_{j}+\frac{1}{2}\log p_{j}\right)-\left(q_{j}+\frac{1}{2}\log q_{j}\right)
\end{equation}

\subsection{Another expression for the deformed logarithm  \cite{kaniadakis2002}.}
It is written as:
\begin{equation}
Log_{\alpha}(x)=\frac{1}{\alpha}\frac{x^{\alpha}-x^{-\alpha}}{x^{\alpha}+x^{-\alpha}}	
\end{equation}
It is defined and concave for $-0.5\leq\alpha\leq 0.5$; it is invariant in the transformation $\alpha\leftrightarrow -\alpha$; the classical logarithm is obtained for $\alpha\rightarrow 0$.\\
By relying on the relation (\ref{entropie classique}) we obtain for the corresponding entropy, the specific expression:
\begin{equation}
	S_{\alpha}(p)=-\sum_{i}p_{i}Log_{\alpha}(p_{i})=-\frac{1}{\alpha}\sum_{i}\frac{p^{\alpha+1}_{i}-p^{-\alpha+1}_{i}}{p^{\alpha}_{i}+p^{-\alpha}_{i}}
\end{equation}
The corresponding basis function, strictly convex for $-0.5\leq\alpha\leq 0.5$, is written:
\begin{equation}
	f(x)=\frac{1}{\alpha}\frac{x^{\alpha+1}-x^{-\alpha+1}}{x^{\alpha}+x^{-\alpha}}
\end{equation}
This is not a standard convex function, because, if $f(1)=0$, we have $f'(1)=1$.\\
The associated strictly convex standard function is therefore written as:
\begin{equation}
	f_{c}(x)=\frac{1}{\alpha}\frac{x^{\alpha+1}-x^{-\alpha+1}}{x^{\alpha}+x^{-\alpha}}-x+1
\end{equation}

\subsubsection{Related Csiszär divergence.}
The Csiszär divergence derived from the previous $f_{c}(x)$ function is written:
\begin{equation}
C_{\alpha}(p\|q)=\frac{1}{\alpha}\sum_{i}\frac{p^{1+\alpha}_{i}q^{-\alpha}_{i}-p^{1-\alpha}_{i}q^{\alpha}_{i}}{p^{\alpha}_{i}q^{-\alpha}_{i}+p^{-\alpha}_{i}q^{\alpha}_{i}}-\sum_{i}p_{i}+\sum_{i}q_{i}
\label{cstri}	
\end{equation}
The calculation of the opposite of the gradient with respect to ``$q$'' leads, after some simple but tedious calculations, to the following expression $\forall j$:
\begin{equation}
-\frac{\partial C_{\alpha}(p\|q)}{\partial q_{j}}=\frac{p_{j}}{q_{j}}\left[\frac{4}{\left(p^{\alpha}_{j}q^{-\alpha}_{j}+p^{-\alpha}_{j}q^{\alpha}_{j}\right)^{2}}\right]-1	
\end{equation}
As one can expect, this gradient is zero if $p_{j}=q_{j}\ \forall j$, moreover the case corresponding to the divergence of Kullback-Leibler is found for $\alpha=0$.

\subsubsection{Related dual Csiszär divergence.}
It is deduced from the expression (\ref{cstri}) and is expressed as:
\begin{equation}
C_{\alpha}(q\|p)=\frac{1}{\alpha}\sum_{i}\frac{q^{1+\alpha}_{i}p^{-\alpha}_{i}-q^{1-\alpha}_{i}p^{\alpha}_{i}}{q^{\alpha}_{i}p^{-\alpha}_{i}+q^{-\alpha}_{i}p^{\alpha}_{i}}-\sum_{i}q_{i}+\sum_{i}p_{i}
\label{cstriduale}	
\end{equation}
The opposite of the gradient is written $\forall j$, after some calculations:
\begin{equation}
-\frac{\partial C_{\alpha}(q\|p)}{\partial q_{j}}=\frac{1}{\alpha}\left(\frac{p^{2a}_{j}-q^{2a}_{j}}{p^{2a}_{j}+q^{2a}_{j}}\right)+\left(\frac{p^{2a}_{j}-q^{2a}_{j}}{p^{2a}_{j}+q^{2a}_{j}}\right)^{2}	
\end{equation}

\subsubsection{Related Bregman and dual Bregman divergences.}
These divergences, which are very heavy to handle, are (perhaps) only of limited interest and will not be developed in this work.

\section{Scale invariant divergences.}
In this section, we develop the divergences invariant by change of scale on ``$q$''.\\
This is because divergences with this property allow a known sum constraint on the corresponding variables to be taken into account in a simple way during the minimisation process.\\
We will develop the calculations in the case of the general divergences (\ref{csgen}) for the Csiszär divergences and (\ref{brgen}) for the Bregman divergences.\\
The specific divergences corresponding to the various entropies will be seen as special cases.\\
However, divergences based on Tsallis entropy represent a very particular case which will be detailed at each step, indeed, for these divergences it is possible to derive a nominal invariance factor whose influence will be examined.\\
We first recall some useful properties of the invariance factors which have been extensively discussed in the references \cite{lanteri2019}, \cite{lanteri2020}.

\subsection{Some recalls on invariance factors and their properties.}
1 - For a divergence $D(p\|q)$, the nominal invariance factor $K_{0}(p,q)$ specific to the divergence considered is obtained as a solution to the equation:
\begin{equation}
	\sum_{i}\frac{\partial D(p_{i}\|Kq_{i})}{\partial K}=0
\end{equation}
if an explicit solution exists.\\\\
2 - The general properties of the invariance factors are as follows:\\\\
	* the invariance factor $K(p,q)$ is a positive scalar.\\
	* The components of the vector $K(p,q).q$ are invariant if ``$q$'' is multiplied by a constant positive factor.\\
	* All the factors $K(p,q)$ having the above properties make any divergence invariant; they are solutions of the differential equation:	
\begin{equation}
	K(p,q)+\sum_{j}q_{j}\frac{\partial K(p,q)}{\partial q_{j}}=0	
\end{equation}
The nominal invariance factors are of course solutions of the above differential equation.\\
	* The general expression for the invariance factors is written as:	
\begin{equation}
	K(p,q)=\left(\frac{\sum_{i}p^{\alpha}_{i}q^{\beta}_{i}}{\sum_{i}p^{\delta}_{i}q^{\gamma}_{i}}\right)^{\mu}	
\end{equation}
With:
\begin{equation}
	\alpha+\beta=\gamma+\delta
\end{equation}
This reflects the fact that $K(p,q)$ is a positive scalar, and:
\begin{equation}
	\mu(\gamma-\beta)=1 
\end{equation}
This reflects the fact that the vector $K(p,q).q$ is invariant with respect to $q$.\\\\
3 - Whether the invariance factor is the nominal invariance factor or not, we have the following fundamental property:
\begin{equation}
\sum_{j}q_{j}\frac{\partial D(p\|K(p,q).q)}{\partial q_{j}}	
\end{equation}

\subsection{Invariant Csiszär divergences.}
We recall that the basic divergence has been written (\ref{csgen}):
\begin{equation}
	C_{ab}(p\|q)=\frac{1}{a-b}\left[\sum_{i}p^{a}_{i}q^{1-a}_{i}-\sum_{i}p^{b}_{i}q^{1-b}_{i}\right]-\sum_{i}p_{i}+\sum_{i}q_{i}
	\label{csgenbis}
\end{equation}
By introducing the invariance factor into this expression, it follows:
\begin{equation}
	C_{ab}(p\|Kq)=\frac{1}{a-b}\left[\sum_{i}K^{1-a}p^{a}_{i}q^{1-a}_{i}-\sum_{i}K^{1-b}p^{b}_{i}q^{1-b}_{i}\right]-\sum_{i}p_{i}+\sum_{i}Kq_{i}
	\label{csgenbisk}
\end{equation} 
The nominal invariance factor ``$K_{0}$'' is obtained as the solution (if it exists) of the equation:
\begin{equation}
	\frac{\partial C_{ab}(p\|Kq)}{\partial K}=0
\end{equation}
This equation is expressed in the present case as follows:
\begin{equation}
	\frac{1-a}{a-b}K^{-a}\sum_{i}p^{a}_{i}q^{1-a}_{i}-\frac{1-b}{a-b}K^{-b}\sum_{i}p^{b}_{i}q^{1-b}_{i}+\sum_{i}q_{i}=0
\end{equation}
In all generality, there is no explicit solution in ``$K$'' except in two particular cases:\\
*1-/ Case of the divergence based on the Tsallis entropy.\\
We then have $a=t$ and $b=1$, which leads to the explicit expression of the nominal invariance factor:
\begin{equation}
	K_{0}=\left(\frac{\sum_{i}p^{t}_{i}q^{1-t}_{i}}{\sum_{i}q_{i}}\right)^{\frac{1}{t}}
	\label{Kzcstsa}
\end{equation}
*2-/ Case of the divergence founded on the Shannon entropy (Kullback-Leibler divergence) which is obtained for example from (\ref{Kzcstsa}) by simply making $t=1$, which leads to:
\begin{equation}
	K=\frac{\sum_{i}p_{i}}{\sum_{i}q_{i}}
	\label{Kgen}
\end{equation}
Consequently, for all the divergences considered, the expression (\ref{Kgen}) (nominal invariance factor for the Kullback-Leibler divergence) will be used as the invariance factor and will serve as a reference.\\
In the case of the divergence based on the Tsallis entropy, we will also develop the invariant divergence using the nominal factor (\ref{Kzcstsa}).

\subsubsection{Csiszär invariant divergences - General case.}
Using the invariance factor (\ref{Kgen}), the divergence (\ref{csgenbisk}) is written, all calculations done, with the notations: $\bar{p}_{i}=\frac{p_{i}}{\sum_{j}p_{j}}$ and $\bar{q}_{i}=\frac{q_{i}}{\sum_{j}q_{j}}$:  
\begin{equation}
	CI_{ab}(p\|q)=\frac{\sum_{j}p_{j}}{a-b}\left[\sum_{i}\overline{p}^{a}_{i}\overline{q}^{1-a}_{i}-\sum_{i}\overline{p}^{b}_{i}\overline{q}^{1-b}_{i}\right]
	\label{csgeninv}
\end{equation}
The opposite of its gradient with respect to ``$q$'' is written $\forall j$, after some calculations:
\begin{align}
-\frac{\partial CI_{ab}(p\|q)}{\partial q_{j}}=\frac{\sum_{j}p_{j}}{\sum_{j}q_{j}}&\left[\frac{a-1}{a-b}\left(\frac{\overline{p}_{j}}{\overline{q}_{j}}\right)^{a}+\frac{1-b}{a-b}\left(\frac{\overline{p}_{j}}{\overline{q}_{j}}\right)^{b}\right. \nonumber \\  & \left.-\frac{a-1}{a-b}\sum_{i}\overline{q}_{i}\left(\frac{\overline{p}_{i}}{\overline{q}_{i}}\right)^{a}-\frac{1-b}{a-b}\sum_{i}\overline{q}_{i}\left(\frac{\overline{p}_{i}}{\overline{q}_{i}}\right)^{b}\right]
\label{mgradcsgeninv}	
\end{align}
From this general form, the expressions corresponding to the entropic divergences given in sections 4 to 9 are easily deduced by making the following adaptations:\\
\begin{itemize}
\item Shannon entropy: $a\rightarrow 1$ and  $b\rightarrow 1$ taking into account (\ref{valab}) or (\ref{valabbis})\\
\item Tsallis entropy: $a=t,$\ \ \ $  b=1$\\
\item Kaniadakis entropy: $a=1+K,$\ \ \ $ b=1-K$\\
\item Abe entropy: $a=z,$\ \ \ $ b=\frac{1}{z}$\\
\item ``$\gamma$'' entropy: $a=2\gamma+1,$\ \ \ $ b=1-\gamma$\\
\item  2 paramèters (KLS) entropy: $a=1+r+K,$\ \ \ $ b=1+r-K$\\
\end{itemize}

\subsubsection{Special case: Csiszär/Tsallis invariant divergence.}
For this divergence, the nominal invariance factor is explicitly computable; its expression is given by (\ref{Kzcstsa}).\\
The invariant divergence obtained using this specific invariance factor is written:
\begin{equation}
	CI_{T}^{K_{0}}(p\|q)=\frac{t}{1-t}\left[\sum_{i}p_{i}-\left(\frac{\sum_{i}p^{t}_{i}q^{1-t}_{i}}{\sum_{i}q_{i}}\right)^{\frac{1}{t}}\sum_{i}q_{i}\right]
	\label{cstsainv}
\end{equation}
Or, more simply:
\begin{equation}
	CI_{T}^{K_{0}}(p\|q)=\frac{t}{1-t}\left[\sum_{i}p_{i}-K_{0}\sum_{i}q_{i}\right]
\end{equation}
The opposite of the gradient with respect to ``$q$'' is written $\forall j$:
\begin{equation}
-\frac{\partial CI_{T}^{K_{0}}(p\|q)}{\partial q_{j}}=\left(\frac{\sum_{i}p^{t}_{i}q^{1-t}_{i}}{\sum_{i}q_{i}}\right)^{\frac{1}{t}-1}	p^{t}_{j}q^{-t}_{j}-\left(\frac{\sum_{i}p^{t}_{i}q^{1-t}_{i}}{\sum_{i}q_{i}}\right)^{\frac{1}{t}}
\label{mgradcstsainvbis}
\end{equation}
By introducing the normalized variables $\overline{p}_{j}=\frac{p_{j}}{\sum_{i}p_{i}}$ and $\overline{q}_{j}=\frac{q_{j}}{\sum_{i}q_{i}}$, we can also write:
\begin{equation}
	-\frac{\partial CI_{T}^{K_{0}}(p\|q)}{\partial q_{j}}=\frac{\sum_{j}p_{j}}{\sum_{j}q_{j}}\left[\left(\sum_{i}\overline{p}^{t}_{i}\overline{q}^{1-t}_{i}\right)^{\frac{1}{t}-1}\overline{p}^{t}_{j}\overline{q}^{-t}_{j}-\left(\sum_{i}\overline{p}^{t}_{i}\overline{q}^{1-t}_{i}\right)^{\frac{1}{t}}\right]
	\label{mgradcstsainv}
\end{equation}

\subsection{Dual invariant Csiszär divergences.}
This divergence is derived from the divergence (\ref{csgenduale})
\begin{equation}
	C_{ab}(q\|p)=\frac{1}{a-b}\left[\sum_{i}q^{a}_{i}p^{1-a}_{i}-\sum_{i}q^{b}_{i}p^{1-b}_{i}\right]-\sum_{i}q_{i}+\sum_{i}p_{i}
	\label{csgendualebis}
\end{equation}
By introducing the invariance factor ``$K$'', it follows:
\begin{equation}
	C_{ab}(Kq\|p)=\frac{1}{a-b}\left[K^{a}\sum_{i}q^{a}_{i}p^{1-a}_{i}-K^{b}\sum_{i}q^{b}_{i}p^{1-b}_{i}\right]-K\sum_{i}q_{i}+\sum_{i}p_{i}
	\label{csgendualebisk}
\end{equation}
The nominal invariance factor is the solution (if it exists) of the equation:
\begin{equation}
\frac{\partial C_{ab}(Kq\|p)}{\partial K}	=0
\label{dpK}
\end{equation}
The expression of this partial derivative is written:
\begin{equation}
\frac{\partial C_{ab}(Kq\|p)}{\partial K}=\frac{1}{a-b}\left[a\;K^{a-1}\sum_{i}q^{a}_{i}p^{1-a}_{i}-b\;K^{b-1}\sum_{i}q^{b}_{i}p^{1-b}_{i}\right]-\sum_{i}q_{i}
\end{equation}
With this expression, the equation (\ref{dpK}) has no general explicit solution in ``$K$''.\\
However, in the particular case of the Tsallis entropy which corresponds to $a=t$ and $b=1$, an explicit solution can be obtained; one then obtain the nominal invariance factor:
\begin{equation}
	K_{0}=\left(\frac{\sum_{j}q_{j}}{\sum_{j}q^{t}_{j}p^{1-t}_{j}}\right)^{\frac{1}{t-1}}
	\label{kztsaduale}
\end{equation}
The nominal invariance factor corresponding to the dual Kullback-Leibler divergence (\ref{divklduale}) can be deduced from this expression by making the passage to the limit $t\rightarrow 1$.\
This leads to the expression which is difficult to exploit:
\begin{equation}
	\log K_{0}=\frac{\sum_{i}q_{i}\log \frac{p_{i}}{q_{i}}}{\sum_{i}q_{i}}
\end{equation}
Consequently, the general case will be developed using the invariance factor:  
\begin{equation}
	K=\frac{\sum_{j}p_{j}}{\sum_{j}q_{j}}
	\label{kgenbis}
\end{equation}
which is the nominal invariance factor for the Kullback-Leibler divergence and we will treat the special case of the Tsallis entropy using its nominal invariance factor given by (\ref{kztsaduale}).

\subsubsection{Dual invariant Csiszär divergences - General case.}
Starting from (\ref{csgendualebisk}), and introducing the invariance factor (\ref{kgenbis}), this divergence is written, with the reduced variables $\overline{q}_{i}=\frac{q_{i}}{\sum_{j}q_{j}}$ and $\overline{p}_{i}=\frac{p_{i}}{\sum_{j}}$:
\begin{equation}
CI_{ab}(q\|p)=\frac{\sum_{j}p_{j}}{a-b}\left(\sum_{i}\overline{q}^{a}_{i}\overline{p}^{1-a}_{i}-\sum_{i}\overline{q}^{b}_{i}\overline{p}^{1-b}_{i}\right)	
\end{equation}
The opposite of the gradient with respect to ``$q$'' is written $\forall j$, all calculations done:
\begin{align}
	-\frac{\partial CI_{ab}(q\|p)}{\partial q_{j}}=\frac{1}{a-b}\frac{\sum_{j}p_{j}}{\sum_{j}q_{j}}&\left[a\ \sum_{i}\overline{q}_{i}\left(\frac{\overline{p}_{i}}{\overline{q}_{i}}\right)^{1-a}-a\ \left(\frac{\overline{p}_{j}}{\overline{q}_{j}}\right)^{1-a}\right. \nonumber \\  & \left.+b\ \sum_{i}\overline{q}_{i}\left(\frac{\overline{p}_{i}}{\overline{q}_{i}}\right)^{1-b}-b\ \left(\frac{\overline{p}_{j}}{\overline{q}_{j}}\right)^{1-b}\right]
\end{align}

\subsubsection{Special case: dual invariant Csiszär/Tsallis divergence.}
In this case, we use the nominal invariance factor given by (\ref{kztsaduale}); from (\ref{csgendualebisk}), we obtain with $a=t$ and $b=1$:
\begin{equation}
	CI_{T}^{K_{0}}(q\|p)=\sum_{i}p_{i}-\left(\frac{\sum_{j}q_{j}}{\sum_{j}q^{t}_{j}p^{1-t}_{j}}\right)^{\frac{1}{t-1}}\sum_{i}q_{i}
\end{equation}
The opposite of its gradient with respect to ``$q$'' is written $\forall j$:
\begin{equation}
-\frac{\partial CI_{T}^{K_{0}}(q\|p)}{\partial q_{j}}=\frac{t}{1-t}\left[\left(\frac{\sum_{j}q_{j}}{\sum_{j}q^{t}_{j}p^{1-t}_{j}}\right)^{\frac{t}{t-1}}\left(\frac{p_{j}}{q_{j}}\right)^{1-t}-\left(\frac{\sum_{j}q_{j}}{\sum_{j}q^{t}_{j}p^{1-t}_{j}}\right)^{\frac{1}{t-1}}\right]	
\end{equation}

\subsection{Invariant Bregman divergences.}
We recall that the Bregman divergence based on the general form of entropy has been written (\ref{brgen}):
\begin{align}
B_{ab}(p\|q)=\frac{1}{a-b}\sum_{i}&\left[p^{a}_{i}-p^{b}_{i}+(a-1)q^{a}_{i}-(b-1)q^{b}_{i}\right. \nonumber \\  & \left.-ap_{i}q^{a-1}_{i}+bp_{i}q^{b-1}_{i}\right]
\label{brgenbis}	
\end{align}
By introducing the invariance factor ``$K$'', it follows:
\begin{align}
B_{ab}(p\|Kq)=\frac{1}{a-b}\sum_{i}&\left[p^{a}_{i}-p^{b}_{i}+(a-1)K^{a}q^{a}_{i}-(b-1)K^{b}q^{b}_{i}\right. \nonumber \\  & \left.-aK^{a-1}p_{i}q^{a-1}_{i}+bK^{b-1}p_{i}q^{b-1}_{i}\right]
\label{brgenbisk}	
\end{align}
The nominal invariance factor is obtained as the solution (if it exists) of the equation:
\begin{equation}
\frac{\partial B_{ab}(p\|Kq)}{\partial K}=0
\label{solK}	
\end{equation}
With:
\begin{align}
\frac{\partial B_{ab}(p\|Kq)}{\partial K}=\frac{1}{a-b}&\left[a(a-1)K^{a-1}\sum_{i}q^{a}_{i}-b(b-1)K^{b-1}\sum_{i}q^{b}_{i}\right. \nonumber \\  & \left.-a(a-1)K^{a-2}\sum_{i}p_{i}q^{a-1}_{i}+b(b-1)K^{b-2}\sum_{i}p_{i}q^{b-1}_{i}\right]	
\end{align}
With this expression, the equation (\ref{solK}) has no explicit solution in all generality.\\
However, in the specific case related to the Tsallis entropy which corresponds to $a=t$ and $b=1$ an explicit solution exists and the nominal invariance factor is written:
\begin{equation}
	K_{0}=\frac{\sum_{i}p_{i}q^{t-1}_{i}}{\sum_{i}q^{t}_{i}}
	\label{Kzbrtsa}
\end{equation}
For $t=1$, we obtain obviously:
\begin{equation}
	K=\frac{\sum_{i}p_{i}}{\sum_{i}q_{i}}
	\label{sboub}
\end{equation}
This is the nominal invariance factor for the Kullback-Leibler divergence based on Shannon entropy.\\
Consequently, for all the divergences considered, we will use this expression as invariance factor; it will serve as our reference.\\
In the case of the divergence based on Tsallis entropy, we will also develop the invariant divergence using the nominal factor (\ref{Kzbrtsa}).

\subsubsection{Bregman invariant divergences - General case.}
By introducing the invariance factor (\ref{sboub}) into the Bregman divergence (\ref{brgenbisk}) based on the general entropy, we obtain all calculations done:
\begin{equation}
\begin{split}
BI_{ab}(p\|q)=& \frac{(\sum_{j}p_{j})^{a}}{a-b} \left[\sum_{i}\overline{p}^{a}_{i}+(a-1)\sum_{i}\overline{q}^{a}_{i}-a\sum_{i}\overline{p}_{i}\overline{q}^{a-1}_{i}\right]\\& -\frac{(\sum_{j}p_{j})^{b}}{a-b}	\left[\sum_{i}\overline{p}^{b}_{i}+(b-1)\sum_{i}\overline{q}^{b}_{i}-b\sum_{i}\overline{p}_{i}\overline{q}^{b-1}_{i}\right]
\end{split}
\label{brgeninv}
\end{equation}
The opposite of its gradient with respect to ``$q$'' is written $\forall j$:
\begin{equation}
\begin{split}
	-\frac{\partial BI_{ab}(p\|q)}{\partial q_{j}}=&\frac{a(a-1)}{(a-b)}\frac{(\sum_{j}p_{j})^{a}}{\sum_{j}q_{j}}\left[\overline{p}_{j}\overline{q}^{a-2}_{j}-\sum_{i}\overline{p}_{i}\overline{q}^{a-1}_{i}-\overline{q}^{a-1}_{j}+\sum_{i}\overline{q}^{a}_{i}\right] \\ & -\frac{b(b-1)}{(a-b)}\frac{(\sum_{j}p_{j})^{b}}{\sum_{j}q_{j}}\left[\overline{p}_{j}\overline{q}^{b-2}_{j}-\sum_{i}\overline{p}_{i}\overline{q}^{b-1}_{i}-\overline{q}^{b-1}_{j}+\sum_{i}\overline{q}^{b}_{i}\right] 
\end{split}
\label{mgradbrgeninv}
\end{equation}
As in the case of the Csiszär divergences, the expressions corresponding to the various entropic divergences given in the previous sections are easily deduced from this general form by making the following adaptations:\\
\begin{itemize}
\item Shannon entropy: $a\rightarrow 1$ and  $b\rightarrow 1$ taking into account (\ref{valab}) or (\ref{valabbis})\\
\item Tsallis entropy: $a=t,$\ \ \ $ b=1$\\
\item Kaniadakis entropy: $a=1+K,$\ \ \ $ b=1-K$\\
\item Abe entropy: $a=z,$\ \ \ $ b=\frac{1}{z}$\\
\item ``$\gamma$'' entropy: $a=2\gamma+1,$\ \ \ $ b=1-\gamma$\\
\item  2 paramèters (KLS) entropy: $a=1+r+K,$\ \ \ $ b=1+r-K$\\
\end{itemize}

\subsubsection{Special case - Bregman/Tsallis invariant divergence.}
In this case, the nominal invariance factor can be calculated explicitly; its expression is given by (\ref{Kzbrtsa}); introducing this factor in (\ref{brgenbisk}), the corresponding invariant divergence is expressed as:
\begin{equation}
	BI_{T}^{K_{0}}(p\|q)=\frac{1}{1-t}\left[\left(\sum_{i}p_{i}q^{t-1}_{i}\right)^{t}\left(\sum_{i}q^{t}_{i}\right)^{1-t}-\sum_{i}p^{t}_{i}\right]
	\label{brtsainv}
\end{equation}
The opposite of its gradient with respect to ``$q$'' is given $\forall j$ by:
\begin{equation}
	-\frac{\partial BI_{T}^{K_{0}}(p\|q)}{\partial q_{j}}=t\left[\left(\frac{\sum_{i}p_{i}q^{t-1}_{i}}{\sum_{i}q^{t}_{i}}\right)^{t-1}p_{j}q^{t-2}_{j}-\left(\frac{\sum_{i}p_{i}q^{t-1}_{i}}{\sum_{i}q^{t}_{i}}\right)^{t}q^{t-1}_{j}\right]
\label{mgradbrtsainv}	
\end{equation}
Or, in a simplified form:
\begin{equation}
	-\frac{\partial BI_{T}^{K_{0}}(p\|q)}{\partial q_{j}}=t\left[K_{0}^{t-1}p_{j}q^{t-2}_{j}-K_{0}^{t}q^{t-1}_{j}\right]
\end{equation}
Or also:
\begin{equation}
	-\frac{\partial BI_{T}^{K_{0}}(p\|q)}{\partial q_{j}}=t\  K_{0}^{t-1}q^{t-1}_{j}\left[\frac{p_{j}}{q_{j}}-K_{0}\right]
\end{equation}

\subsection{Dual Bregman divergences - invariant form.}
We will use the form of this divergence written in (\ref{brgenduale}):
\begin{align}
B_{ab}(q\|p)=\frac{1}{a-b}\sum_{i}&\left[q^{a}_{i}-q^{b}_{i}+(a-1)p^{a}_{i}-(b-1)p^{b}_{i}\right. \nonumber \\  & \left.-aq_{i}p^{a-1}_{i}+bq_{i}p^{b-1}_{i}\right]
\label{brgendualebis}	
\end{align}
By introducing the invariance factor ``$K$'', it follows:
\begin{align}
B_{ab}(Kq\|p)=\frac{1}{a-b}\sum_{i}&\left[K^{a}q^{a}_{i}-K^{b}q^{b}_{i}+(a-1)p^{a}_{i}-(b-1)p^{b}_{i}\right. \nonumber \\  & \left.-aKq_{i}p^{a-1}_{i}+bKq_{i}p^{b-1}_{i}\right]
\label{brgendualek}	
\end{align}
The nominal invariance factor is obtained as the solution (if it exists) of the equation:
\begin{equation}
\frac{\partial B_{ab}(Kq\|p)}{\partial K}=0
\label{solKbis}	
\end{equation}
With:
\begin{align}
\frac{\partial B_{ab}(Kq\|p)}{\partial K}=\frac{1}{a-b}&\left[aK^{a-1}\sum_{i}q^{a}_{i}-bK^{b-1}\sum_{i}q^{b}_{i}\right. \nonumber \\  & \left.-a\sum_{i}q_{i}p^{a-1}_{i}+b\sum_{i}q_{i}p^{b-1}_{i}\right]	
\end{align}
With this expression, the equation (\ref{solKbis}) has no explicit solution in all generality.\\
However, in the specific case related to the Tsallis entropy which corresponds to  $a=t$ and $b=1$ an explicit solution exists and the nominal invariance factor is written:
\begin{equation}
	K_{0}=\left(\frac{\sum_{i}q^{t}_{i}}{\sum_{i}q_{i}p^{t-1}_{i}}\right)^{\frac{1}{1-t}}
	\label{Kzbrtsaduale}
\end{equation}
Consequently, for all the divergences considered, we will use as an invariance factor the expression:
\begin{equation}
	K=\frac{\sum_{i}p_{i}}{\sum_{i}q_{i}}
	\label{sboubbis}
\end{equation}
This is the nominal invariance factor for the Kullback-Leibler divergence based on Shannon entropy; it will serve as our reference.\\
In the case of the divergence based on Tsallis entropy, we will also develop the invariant divergence using the nominal factor (\ref{Kzbrtsaduale}).

\subsubsection{Dual invariant Bregman divergences - General case.}
By introducing the invariance factor (\ref{sboubbis}) into the Bregman divergence (\ref{brgendualek}) based on the general entropy, we obtain all calculations done:
\begin{equation}
\begin{split}
	BI_{ab}(q\|p)=&\frac{(\sum_{j}p_{j})^{a}}{a-b}\left[\sum_{i}\overline{q}^{a}_{i}+(a-1)\sum_{i}\overline{p}^{a}_{i}-a\sum_{i}\overline{q}_{i}\overline{p}^{(a-1)}_{i}\right]\\&-\frac{(\sum_{j}p_{j})^{b}}{a-b}\left[\sum_{i}\overline{q}^{b}_{i}+(b-1)\sum_{i}\overline{p}^{b}_{i}-b\sum_{i}\overline{q}_{i}\overline{p}^{(b-1)}_{i}\right]
\end{split}
\end{equation}
The opposite of its gradient with respect to ``$q$'' is written $\forall j$, after some calculations:
\begin{equation}
\begin{split}
	-\frac{\partial BI_{ab}(q\|p)}{\partial q_{j}}=&\frac{b\ (\sum_{j}p_{j})^{b}}{(a-b)\sum_{j}q_{j}}\left(\overline{q}^{(b-1)}_{j}-\sum_{i}\overline{q}^{b}_{i}-\overline{p}^{(b-1)}_{j}+\sum_{i}\overline{q}_{i}\overline{p}^{(b-1)}_{i}\right)\\&-\frac{a\ (\sum_{j}p_{j})^{a}}{(a-b)\sum_{j}q_{j}}\left(\overline{q}^{(a-1)}_{j}-\sum_{i}\overline{q}^{a}_{i}-\overline{p}^{(a-1)}_{j}+\sum_{i}\overline{q}_{i}\overline{p}^{(a-1)}_{i}\right)
\end{split}
\end{equation}
The expressions corresponding to the different entropies are obtained, as indicated in the previous sections, by affecting to ``$a$'' and ``$b$'', the values of the parameters specific to each case.

\subsubsection{Special case of the dual invariant Bregman/Tsallis divergence.}
In this case, the nominal invariance factor is computed explicitly; it is given by the relation (\ref{Kzbrtsaduale}).\\
By introducing it in the expression (\ref{brgendualek}), we obtain the invariant divergence:
\begin{equation}
	BI_{T}^{K_{0}}(q\|p)=\sum_{i}p^{t}_{i}-\left(\frac{\sum_{i}q^{t}_{i}}{\sum_{i}q_{i}p^{t-1}_{i}}\right)^{\frac{t}{1-t}}\sum_{i}q^{t}_{i}
\end{equation}
Which can be written more simply:
\begin{equation}
	BI_{T}^{K_{0}}(q\|p)=\sum_{i}p^{t}_{i}-K^t_{0}\sum_{i}q^{t}_{i}
\end{equation}
The opposite of the gradient with respect to ``$q$'' is written $\forall j$:
\begin{equation}
	-\frac{\partial BI_{T}^{K_{0}}(q\|p)}{\partial q_{j}}=\frac{t}{t-1}\left[\left(\frac{\sum_{i}q^{t}_{i}}{\sum_{i}q_{i}p^{1-t}_{i}}\right)^{\frac{1}{1-t}}p^{t-1}_{j}-\left(\frac{\sum_{i}q^{t}_{i}}{\sum_{i}q_{i}p^{1-t}_{i}}\right)^{\frac{t}{1-t}}q^{t-1}_{j}\right]
\end{equation}
Or else, in another form:
\begin{equation}
	-\frac{\partial BI_{T}^{K_{0}}(q\|p)}{\partial q_{j}}=\frac{t}{t-1}K_{0}\left[p^{t-1}_{j}-K^{t-1}_{0}q^{t-1}_{j}\right]
\end{equation}

\section{Algorithmic aspect}
\subsection{General informations on the method used to build the algorithms.}
We recall that in the case we are concerned with, we have $p=y$ and $q=Hx$.\\
Then, for a divergence $D(p\|q)$ or $D(q\|p)$, we have:
\begin{equation}
-\frac{\partial D}{\partial x}=-H^{T}\frac{\partial D}{\partial q}	
\end{equation}
This justifies the fact that throughout this work, the expressions (of the opposite) of the gradients with respect to $q$ have been given.\\
The algorithms proposed here are based on the SGM method or its variants explained in detail in \cite{lanteri2019} and \cite{lanteri2020}.\\
If the divergence considered is of classical non-invariant form, the proposed algorithms take into account the non-negativity constraint of the solution.\\
On the other hand, if one wants also to take into account a constraint of sum on the components of the solution, the invariant divergences must be used.\\ 
In all cases, the general form of the algorithms remains the same.\\
The basic iterative algorithm is written in general form:
\begin{equation}
x^{k+1}_{l}=x^{k}_{l}+\alpha^{k}_{l}x^{k}_{l}\left(-\frac{\partial D}{\partial x}\right)^{k}_{l}	
\end{equation}
The divergence $D$ being strictly convex with respect to ``$x$'', $\left(-\frac{\partial D}{\partial x}\right)$ is a direction of descent.\\
The opposite of the gradient can always be written as:
\begin{equation}
\left(-\frac{\partial D}{\partial x}\right)^{k}_{l}=U^{k}_{l}-V^{k}_{l}\ \ ;\ \ \ \ U^{k}_{l}>0\ \ ;\ \ \ \ V^{k}_{l}>0	
\end{equation}
Then:
\begin{equation}
x^{k+1}_{l}=x^{k}_{l}+\alpha^{k}_{l}x^{k}_{l}\left(U^{k}_{l}-V^{k}_{l}\right)	
\end{equation}
And with preconditioning by $\frac{1}{V^{k}_{l}}>0$:
\begin{equation}
x^{k+1}_{l}=x^{k}_{l}+\alpha^{k}_{l}x^{k}_{l}\left(\frac{U^{k}_{l}}{V^{k}_{l}}-1\right)	
\end{equation}
Given $V^{k}_{l}>0$, the modified negative gradient $\frac{U^{k}}{V^{k}}-1$ remains a descent direction.\\\\
\textbf{Comment on the notations}: in this writing, $\frac{U^{k}}{V^{k}}$ is a vector obtained by taking the component-to-component ratio of the vectors $U^{^{k}}$ and $V^{^{k}}$, and ``$1$'' is the unit vector.\\\\ 
Regarding the descent stepsize, for each of these two algorithms, we adopt the following procedure:\\
* 1 - At a given iteration, we compute the maximum stepsize $(\alpha^{k})_{Max}$ that ensures the non-negativity of the set of components of $x^{k+1}$.\\
* 2 - The descent stepsize $\alpha^{k}$ (valid for all the components), which ensures the convergence of the algorithm is then computed by a one-dimensional search method of Armijo type (for example), in the interval $\left[0,(\alpha^{k})_{Max}\right]$.\\
This procedure is described in detail in \cite{lanteri2019} and \cite{lanteri2020}.\\
We can write in a general way, with a descent step independent of the component:
\begin{equation}
x^{k+1}=x^{k}+\alpha^{k}x^{k}\left(U^{k}-V^{k}\right)
\label{algobase}	
\end{equation}
\textbf{Comment on the notations}: in this expression, the operation $x^{k}\left(U^{k}-V^{k}\right)$ represents the component-to-component product of the vectors $x^{k}$ and $\left(U^{k}-V^{k}\right)$ (Hadamard product).\\\\
With a modified (preconditioned) gradient, we have:
\begin{equation}
x^{k+1}=x^{k}+\alpha^{k}x^{k}\left(\frac{U^{k}}{V^{k}}-1\right)
\label{algoprecond}	
\end{equation}
In the latter case, if we use a descent step $\alpha^{k}=1\ ,\ \forall k$, we obtain a purely multiplicative algorithm which is written
\begin{equation}
x^{k+1}=x^{k}\left(\frac{U^{k}}{V^{k}}\right)
\label{algomult}	
\end{equation}
Of course, in all generality, nothing proves the convergence of purely multiplicative algorithms, each divergence implies a particular analysis.\\\\
\textbf{Note: The comments concerning the notations are valid for all the algorithms proposed in the following sections, more precisely, operations between vectors (products or ratios), are componentwise operations; the result is always a vector.}\\\\

\textbf{Remarks}:\\
* 1 - For non-invariant divergences, the algorithms (\ref{algobase}) (\ref{algoprecond}) and (\ref{algomult}) only ensure the non-negativity of the solution.\\
* 2 - If one requires in addition that the sum constraint is fulfilled, the invariant divergences take all their significance, indeed, an algorithm of type (\ref{algobase}) makes it possible to ensure the property:
\begin{equation}
	\sum_{l}x^{k+1}_{l}=\sum_{l}x^{k}_{l}
\end{equation}
Starting from an initial estimate $x^{0}$ such that $\sum_{l}x^{0}_{l}=C$, all the successive estimates will have the same sum.\\
On the other hand, the use of a preconditionned algorithm (\ref{algoprecond}) or of a purely multiplicative algorithm of the type (\ref{algomult}) (as long as its convergence is ensured), does not automatically ensure the sum constraint; an additional step is necessary:\\
At each iteration, the operation is performed in 2 steps:\\\\
* - First, a temporary estimate is calculated:
\begin{equation}
\widetilde{x}^{k+1}=x^{k}\left(\frac{U^{k}}{V^{k}}\right)	
\end{equation}
* - Then in a normalisation step, we compute:
\begin{equation}
	x^{k+1}=\frac{\widetilde{x}^{k+1}}{\sum_{l}\tilde{x}^{k+1}_{l}}C
\end{equation}
Considering the properties of invariant divergences, this last operation does not modify the value of the divergence concerned.

\subsection{Algorithms based on general entropy.}
We describe in this section the algorithms based on the forms of divergences related to the general expression of the entropy.\\
We recall that the domains of values of the parameters ``$a$'' and ``$b$'' appearing in these divergences are given by (\ref{valab}) (\ref{valabbis}):
\begin{equation}
	0\leq a\leq 1\leq b
\end{equation}
or else:
\begin{equation}
 0\leq b\leq 1\leq a
\end{equation}
Algorithms coresponding to dual divergences based on general entropy, whether invariant or non-invariant, will not be detailed here, but this does not present any particular difficulty apart from the specific problems already mentioned.
 
\subsubsection{Csiszär divergence based on the general entropy.}
The expression of this divergence is given by (\ref{csgen}); considering the expression of the opposite of the gradient given in (\ref{mgradcsgen}), the algorithm (\ref{algobase}) is written here: 
\begin{equation}
		x^{k+1}=x^{k}+\alpha^{k}x^{k}H^{T}\left\{\underbrace{\left[\frac{a-1}{a-b}\left(\frac{p}{q^{k}}\right)^{a}+\frac{1-b}{a-b}\left(\frac{p}{q^{k}}\right)^{b}\right]}_{A^{k}}-\underbrace{1}_{B^{k}}\right\}
\end{equation}
The corresponding multiplicative form is immediately derived by (\ref{algomult}) using the expressions: 
\begin{equation}
	U^{k}=H^{T}A^{k}\ \ \ ;\ \ \ V^{k}=H^{T}B^{k}
\end{equation}

\subsubsection{Bregman divergence based on the general entropy.}
The expression of this divergence is given by (\ref{brgen}); considering the expression of the opposite of the gradient given in (\ref{mgradbrgen}), the algorithm (\ref{algobase}) is written here:
\begin{equation}
	x^{k+1}=x^{k}+\alpha^{k}x^{k}H^{T}\left\{\left(\frac{p}{q^{k}}-1\right)\left[\frac{(a^{2}-a)}{a-b}(q^{k})^{a-1}+\frac{(b-b^{2})}{a-b}(q^{k})^{b-1}\right]\right\}
\end{equation}
The corresponding multiplicative form is immediately obtained by (\ref{algomult}), with:
\begin{equation}
	U^{k}=H^{T}\left\{\left(\frac{p}{q^{k}}\right)\left[\frac{(a^{2}-a)}{a-b}(q^{k})^{a-1}+\frac{(b-b^{2})}{a-b}(q^{k})^{b-1}\right]\right\}
\end{equation}
and
\begin{equation}
	V^{k}=H^{T}\left[\frac{(a^{2}-a)}{a-b}(q^{k})^{a-1}+\frac{(b-b^{2})}{a-b}(q^{k})^{b-1}\right]
\end{equation}

\subsubsection{Csiszär invariant divergence based on the general entropy.}
The expression of this divergence is given by (\ref{csgeninv}), and the opposite of its gradient is given by (\ref{mgradcsgeninv}); the algorithm (\ref{algobase}) is written here: 
\begin{align}
x^{k+1}=x^{k}+\alpha^{k}x^{k}H^{T}\frac{\sum_{j}p_{j}}{\sum_{j}q^{k}_{j}}&\left[\frac{a-1}{a-b}\left(\frac{\overline{p}}{\overline{q}^{k}}\right)^{a}+\frac{1-b}{a-b}\left(\frac{\overline{p}}{\overline{q}^{k}}\right)^{b}\right. \nonumber \\  & \left.-\frac{a-1}{a-b}\sum_{i}\overline{q}^{k}_{i}\left(\frac{\overline{p}_{i}}{\overline{q}^{k}_{i}}\right)^{a}-\frac{1-b}{a-b}\sum_{i}\overline{q}^{k}_{i}\left(\frac{\overline{p}_{i}}{\overline{q}^{k}_{i}}\right)^{b}\right]
\label{algocsgeninv}	
\end{align}
The decomposition allowing to obtain a multiplicative algorithm is written:
\begin{equation}
	U^{k}=H^{T}\frac{\sum_{j}p_{j}}{\sum_{j}q^{k}_{j}}\left[\frac{a-1}{a-b}\left(\frac{\overline{p}}{\overline{q}^{k}}\right)^{a}+\frac{1-b}{a-b}\left(\frac{\overline{p}}{\overline{q}^{k}}\right)^{b}\right]
\end{equation}
and
\begin{equation}
	V^{k}=H^{T}\frac{\sum_{j}p_{j}}{\sum_{j}q^{k}_{j}}\left[\frac{a-1}{a-b}\sum_{i}\overline{q}^{k}_{i}\left(\frac{\overline{p}_{i}}{\overline{q}^{k}_{i}}\right)^{a}+\frac{1-b}{a-b}\sum_{i}\overline{q}^{k}_{i}\left(\frac{\overline{p}_{i}}{\overline{q}^{k}_{i}}\right)^{b}\right]
\end{equation}\\

\textbf{* Special case of the Csiszär/Tsallis invariant divergence.}\\
Taking into account the fact that for this type of entropy, the nominal invariance factor can be calculated explicitly and leads to a simplified algorithm, we detail here, this specific case.\\
We recall that the expression of the nominal invariance factor is given by the relation (\ref{Kzcstsa}).\\
This leads to the invariant divergence (\ref{cstsainv}) whose opposite of the gradient is given by (\ref{mgradcstsainv}).\\
The algorithm (\ref{algobase}) is written accordingly:
\begin{equation}
x^{k+1}=x^{k}+\alpha^{k}x^{k}H^{T}\frac{\sum_{j}p_{j}}{\sum_{j}q^{k}_{j}}\left(\sum_{i}\overline{p}^{t}_{i}(\overline{q}^{k}_{i})^{1-t}\right)^{\frac{1}{t}-1}\left[\left(\frac{\overline{p}}{\overline{q}^{k}}\right)^{t}-\sum_{i}\overline{q}^{k}_{i}\left(\frac{\overline{p}_{i}}{\overline{q}^{k}_{i}}\right)^{t}\right]
\end{equation}
The decomposition allowing to write a multiplicative algorithm is obvious.\\
We give below, the analogous algorithm deduced from (\ref{algocsgeninv}) by making $a=t$ and $b=1$.\\
This algorithm corresponds to a non-nominal invariance factor for the Tsallis entropy, it is written:
\begin{equation}
x^{k+1}=x^{k}+\alpha^{k}x^{k}H^{T}\frac{\sum_{j}p_{j}}{\sum_{j}q^{k}_{j}}\left[\left(\frac{\overline{p}}{\overline{q}^{k}}\right)^{t}-\sum_{i}\overline{q}^{k}_{i}\left(\frac{\overline{p}_{i}}{\overline{q}^{k}_{i}}\right)^{t}\right]
\end{equation}

\subsubsection{Bregman invariant divergence based on the general entropy.}
The expression of this divergence is given by (\ref{brgeninv}), and the opposite of its gradient is given by (\ref{mgradbrgeninv}); the algorithm (\ref{algobase}) is written here:
\begin{align}
	x^{k+1}=x^{k}+&\alpha^{k}x^{k}H^{T}\left[\frac{a(a-1)}{(a-b)}\frac{(\sum_{j}p_{j})^{a}}{\sum_{j}q^{k}_{j}}\left(\overline{p}(\overline{q}^k)^{{a-2}}-\sum_{i}\overline{p}_{i}(\overline{q}^k_{i})^{a-1}-(\overline{q}^k)^{a-1}+\sum_{i}(\overline{q}^k_{i})^{a}\right)\right. \nonumber \\  & \left.-\frac{b(b-1)}{(a-b)}\frac{(\sum_{j}p_{j})^{b}}{\sum_{j}q^{k}_{j}}\left(\overline{p}(\overline{q}^k)^{b-2}-\sum_{i}\overline{p}_{i}(\overline{q}^k_{i})^{b-1}-(\overline{q}^k)^{b-1}+\sum_{i}(\overline{q}^k_{i})^{b}\right)\right]
	\label{algobrgeninv}
\end{align}
The decomposition allowing to write a multiplicative algorithm is as follows:
\begin{align}
	U^{k}=H^{T}&\left[\frac{a(a-1)}{(a-b)}\frac{(\sum_{j}p_{j})^{a}}{\sum_{j}q^{k}_{j}}\left(\overline{p}(\overline{q}^k)^{{a-2}}+\sum_{i}(\overline{q}^k_{i})^{a}\right)\right. \nonumber \\  & \left.+\frac{b(1-b)}{(a-b)}\frac{(\sum_{j}p_{j})^{b}}{\sum_{j}q^{k}_{j}}\left(\overline{p}(\overline{q}^k)^{b-2}+\sum_{i}(\overline{q}^k_{i})^{b}\right)\right] 
\end{align}
\begin{align}
	V^{k}=H^{T}&\left[\frac{a(a-1)}{(a-b)}\frac{(\sum_{j}p_{j})^{a}}{\sum_{j}q^{k}_{j}}\left(\sum_{i}\overline{p}_{i}(\overline{q}^k_{i})^{a-1}+(\overline{q}^k)^{a-1}\right)\right. \nonumber \\  & \left.+\frac{b(1-b)}{(a-b)}\frac{(\sum_{j}p_{j})^{b}}{\sum_{j}q^{k}_{j}}\left(\sum_{i}\overline{p}_{i}(\overline{q}^k_{i})^{b-1}+(\overline{q}^k)^{b-1}\right)\right] 
\end{align}\\

\textbf{* Special case of the Bregman/Tsallis invariant divergence.}\\
Taking into account the fact that for this type of entropy, the nominal invariance factor can be calculated explicitly and leads to a simplified algorithm, we detail here, this specific case.\\
We recall that the expression of the nominal invariance factor is given by the relation (\ref{Kzbrtsa}).\\
This leads to the invariant divergence (\ref{brtsainv}) whose opposite of the gradient is given by (\ref{mgradbrtsainv}).\\
The corresponding algorithm (\ref{algobase}) is written accordingly:
\begin{align}
	x^{k+1}=x^{k}+\alpha^{k}x^{k}H^{T}\ t\ &\left[\left(\frac{\sum_{i}p_{i}(q^{k}_{i})^{t-1}}{\sum_{i}(q^{k}_{i})^{t}}\right)^{t-1}p\ (q^{k})^{t-2}\right. \nonumber \\  & \left.-\left(\frac{\sum_{i}p_{i}(q^{k}_{i})^{t-1}}{\sum_{i}(q^{k}_{i})^{t}}\right)^{t}(q^{k})^{t-1}\right]
\end{align}
Note that the term $(p\ q^{t-2})$ is a vector obtained by taking the product component wise of the vectors $p$ and $q^{t-2}$.
As a comparison, we explicit hereafter, the analogous algorithm deduced from the expression (\ref{algobrgeninv}) by making $a=t$ and $b=1$ and which corresponds to a non-nominal invariance factor for the Tsallis entropy.
\begin{align}
	x^{k+1}=x^{k}+\alpha^{k}x^{k}H^{T}\ t\ \frac{\sum_{j}(p_{j})^{t}}{\sum_{j}q^{k}_{j}}&\left[\overline{p}(\overline{q}^k)^{t-2}-\sum_{i}\overline{p}_{i}(\overline{q}^k_{i})^{t-1}\right. \nonumber \\  & \left.-(\overline{q}^k)^{t-1}+\sum_{i}(\overline{q}^k_{i})^{t}\right]
\end{align}

\section{Appendix.}
We summarise here the expressions of the opposite of the gradients $\left(-\frac{\partial D}{\partial q_{j}}\right)$ for the various divergences examined in the previous sections, with the aim of highlighting specific synthetic forms of these expressions.

\subsection{Csiszär divergences.}
These expressions can be written $\forall j$ in synthetic form:
\begin{equation}
\frac{p_{j}}{q_{j}}X_{j}-1
\end{equation}
With for $X_{j}$, the following expressions:\\
\begin{itemize}
\item Shannon entropy:$\ \ X_{j}=1$\\
\item Tsallis entropy:$\ \ X_{j}=\left(\frac{p_{j}}{q_{j}}\right)^{t-1}$\\ 
\item Kaniadakis entropy:$\ \ X_{j}=\frac{1}{2}\left(\frac{p_{j}}{q_{j}}\right)^{K}+\frac{1}{2}\left(\frac{p_{j}}{q_{j}}\right)^{-K}$\\ 
\item Abe entropy:$\ \ \ \ X_{j}=\frac{z}{z+1}\left(\frac{p_{j}}{q_{j}}\right)^{z-1}+\frac{1}{z+1}\left(\frac{p_{j}}{q_{j}}\right)^{\frac{1}{z}-1}$\\
\item ``$\gamma$'' entropy:$\ \ X_{j}=\frac{2}{3}\left(\frac{p_{j}}{q_{j}}\right)^{2\gamma}+\frac{1}{3}\left(\frac{p_{j}}{q_{j}}\right)^{-\gamma}$\\
\item 2 parameters (KLS) entropy:$\ \ X_{j}=\frac{K+r}{2K}\left(\frac{p_{j}}{q_{j}}\right)^{r+K}+\frac{K-r}{2K}\left(\frac{p_{j}}{q_{j}}\right)^{r-K}$\\
\item General entropy:$\ \ X_{j}=\frac{a-1}{a-b}\left(\frac{p_{j}}{q_{j}}\right)^{a-1}+\frac{1-b}{a-b}\left(\frac{p_{j}}{q_{j}}\right)^{b-1}$\\
\item Newton entropy:$\ \ X_{j}=\frac{1}{2}\left(\frac{p_{j}}{q_{j}}\right)+\frac{1}{2}$\\
\end{itemize}

\subsection{Dual Csiszär divergences.}
In order to have a synthetic writing for the expressions of the opposite of the gradients, we first make the following writing changes (which may seem a bit tortuous):\\
For the Shannon entropy:\\\\
$\log\frac{p_{j}}{q_{j}}=\left[\log \frac{p_{j}}{q_{j}}-1\right]+1=\frac{p_{j}}{q_{j}}\left[\left(\frac{p_{j}}{q_{j}}\right)^{-1}\log \frac{p_{j}}{q_{j}}-\left(\frac{p_{j}}{q_{j}}\right)^{-1}\right]+1$\\\\
For the Tsallis entropy:\\\\
$\left[\frac{t}{1-t}\left(\frac{p_{j}}{q_{j}}\right)^{1-t}-\frac{1}{1-t}\right]+1=\frac{p_{j}}{q_{j}}\left[\frac{t}{1-t}\left(\frac{p_{j}}{q_{j}}\right)^{-t}-\frac{1}{1-t}\left(\frac{p_{j}}{q_{j}}\right)^{-1}\right]+1$\\\\
Then the expressions for the opposite of the gradients can be written $\forall j$ in the synthetic form:
\begin{equation}
	\frac{p_{j}}{q_{j}}\ T_{j}+1
\end{equation}
With for $T_{j}$, the following expressions:\\
\begin{itemize}
\item Shannon entropy:$\ \ T_{j}=\left(\frac{p_{j}}{q_{j}}\right)^{-1}\log \frac{p_{j}}{q_{j}}-\left(\frac{p_{j}}{q_{j}}\right)^{-1}$\\
\item Tsallis entropy:$\ \ T_{j}=\frac{t}{1-t}\left(\frac{p_{j}}{q_{j}}\right)^{-t}-\frac{1}{1-t}\left(\frac{p_{j}}{q_{j}}\right)^{-1}$\\
\item Kaniadakis entropy:$\ \ T_{j}=\frac{1-K}{2K}\left(\frac{p_{j}}{q_{j}}\right)^{K-1}-\frac{1+K}{2K}\left(\frac{p_{j}}{q_{j}}\right)^{-K-1}$\\
\item Abe entropy:$\ \ T_{j}=\frac{z^{2}}{1-z^{2}}\left(\frac{p_{j}}{q_{j}}\right)^{-z}-\frac{1}{1-z^{2}}\left(\frac{p_{j}}{q_{j}}\right)^{-\frac{1}{z}}$\\
\item ``$\gamma$'' entropy:$\ \ T_{j}=\frac{1-\gamma}{3\gamma}\left(\frac{p_{j}}{q_{j}}\right)^{\gamma-1}-\frac{1+2\gamma}{3\gamma}\left(\frac{p_{j}}{q_{j}}\right)^{-2\gamma-1}$\\
\item 2 parameters entropy:$\ \ T_{j}=\frac{1+r-K}{2K}\left(\frac{p_{j}}{q_{j}}\right)^{K-r-1}-\frac{1+r+K}{2K}\left(\frac{p_{j}}{q_{j}}\right)^{-K-r-1}$\\
\item General entropy:$\ \ T_{j}=\frac{b}{a-b}\left(\frac{p_{j}}{q_{j}}\right)^{-b}-\frac{a}{a-b}\left(\frac{p_{j}}{q_{j}}\right)^{-a}$\\
\item Newton entropy:$\ \ T_{j}=\frac{1}{2}\left(\frac{p_{j}}{q_{j}}\right)^{-1}\log\left(\frac{p_{j}}{q_{j}}\right)-1$\\
\end{itemize}

\subsection{Bregman divergences.}
These expressions can be written $\forall j$ in the synthetic form:\\
\begin{equation}
\left[\frac{p_{j}}{q_{j}}-1\right]\ Z_{j}	
\end{equation}
With for $Z_{j}$, the following expressions:\\

\begin{itemize}
\item Shannon entropy:$\ \ Z_{j}=1$\\
\item Tsallis entropy:$\ \ Z_{j}=t\ q^{t-1}_{j}$\\
\item Kaniadakis entropy:$\ \ Z_{j}=\frac{1+K}{2}q^{K}_{j}+\frac{1-K}{2}q^{-K}_{j}$\\
\item Abe entropy:$\ \ Z_{j}=\frac{z^{2}}{z+1}q^{z-1}_{j}+\frac{1}{z(z+1)}q^{\frac{1}{z}-1}_{j}$\\
\item ``$\gamma$'' entropy:$\ \ Z_{j}=\frac{2(2\gamma+1)}{3}q^{2\gamma}_{j}+\frac{1-\gamma}{3}q^{-\gamma}_{j}$\\
\item 2 parameters entropy:$\ \ Z_{j}=\frac{(r+K)(1+r+K)}{2K}q^{r+K}_{j}+\frac{(r-K)(1+r-K)}{2K}q^{r-K}_{j}$\\
\item General entropy:$\ \ Z_{j}=\frac{a^{2}-a}{a-b}q^{a-1}_{j}+\frac{b-b^{2}}{a-b}q^{b-1}_{j}$\\
\item Newton entropy:$\ \ Z_{j}=\frac{1}{2}+q_{j}$\\
\end{itemize}

\subsection{Dual Bregman divergences.}
A general synthetic form is not easy to write, even if one observes a clear analogy between these expressions
that can be written $\forall j$:\\
\begin{itemize}
\item Shannon entropy:$\ \ \log\frac{p_{j}}{q_{j}}=\log p_{j}-\log q_{j}$\\
\item Tsallis entropy:$\ \ \frac{t}{1-t}\left[p^{t-1}_{j}-q^{t-1}_{j}\right]$\\
\item Kaniadakis entropy:$\ \ \frac{1+K}{2K}(p^{K}_{j}-q^{K}_{j})-\frac{1-K}{2K}(p^{-K}_{j}-q^{-K}_{j})$\\
\item Abe entropy:$\ \ \frac{z^{2}}{z^{2}-1}(p^{z-1}_{j}-q^{z-1}_{j})-\frac{1}{z^{2}-1}(p^{\frac{1}{z}-1}_{j}-q^{\frac{1}{z}-1}_{j})$\\
\item ``$\gamma$'' entropy:$\ \ \frac{2\gamma+1}{3\gamma}\left(p^{2\gamma}_{j}-q^{2\gamma}_{j}\right)-\frac{1-\gamma}{3\gamma}\left(p^{-\gamma}_{j}-q^{-\gamma}_{j}\right)$\\
\item 2 parameters entropy:$\ \ \frac{1+r+K}{2K}\left(p^{r+K}_{j}-q^{r+K}_{j}\right)-\frac{1+r-K}{2K}\left(p^{r-K}_{j}-q^{r-K}_{j}\right)$\\
\item General entropy:$\ \ \frac{a}{a-b}\left(p^{a-1}_{j}-q^{a-1}_{j}\right)-\frac{b}{a-b}\left(p^{b-1}_{j}-q^{b-1}_{j}\right)$\\
\item Newton entropy:$\ \ \ (p_{j}-q_{j})+\frac{1}{2}\log\frac{p_{j}}{q_{j}}=\left(p_{j}+\frac{1}{2}\log p_{j}\right)-\left(q_{j}+\frac{1}{2}\log q_{j}\right)$\\
\end{itemize}

\
\bibliographystyle{plain}
\bibliography{biblio}

\end{document}